\documentclass[12pt]{article}
\usepackage{amsfonts}
\usepackage{mathrsfs}
\usepackage{amsthm}
\usepackage{multirow}
\usepackage{bbding}
\usepackage{amssymb}
\usepackage{amsmath}
\usepackage{graphicx,color}
\usepackage{rotating}

\usepackage{bm}
\usepackage{epstopdf}
\newcommand{\bea}{\begin{eqnarray}}
\newcommand{\eea}{\end{eqnarray}}
\newcommand{\Bea}{\begin{eqnarray*}}
\newcommand{\Eea}{\end{eqnarray*}}
\newcommand{\ba}{\begin{array}}
\newcommand{\ea}{\end{array}}
\newcommand{\bt}{\begin{tabular}}
\newcommand{\et}{\end{tabular}}
\newcommand{\btb}{\begin{table}}
\newcommand{\etb}{\end{table}}
\newcommand{\bc}{\begin{center}}
\newcommand{\ec}{\end{center}}
\newcommand{\beq}{\begin{equation}}
\newcommand{\eeq}{\end{equation}}

\usepackage{float}

\newtheorem{theorem}{\sc Theorem}[section]

\newtheorem{coro}{\sc Corollary}[section]
\newtheorem{rema}{\sc Remark}[section]

\makeatletter

\newcommand{\Rmnum}[1]{\expandafter\@slowromancap\romannumeral #1@}
\makeatother

\begin{document}

\title{Online Updating Statistics for Heterogenous Updating Regressions via Homogenization Techniques}
\author{Lu Lin$^1$, Jun Lu$^2$\footnote{The corresponding
author. Email: jlustat@gmail.com. The research was
supported by NNSF project (11971265, 12001486) of China.} \ and Weiyu Li$^1$
\\
\small $^1$Zhongtai Securities Institute for Financial Studies, Shandong University, Jinan, China\\
\small $^2$School of Statistics and Mathematics, Zhejiang Gongshang University, Hangzhou, China
}
\date{}
\maketitle

\vspace{-0.3cm}

\begin{abstract} \baselineskip=18pt
Under the environment of big data streams, it is a common situation where the variable set of a model may change according to the condition of data streams. In this paper, we propose a homogenization strategy to represent the heterogenous models that are  gradually updated in the process of data streams. With the homogenized representations, we can easily construct various online updating statistics such as parameter estimation, residual sum of squares and $F$-statistic for the heterogenous updating regression models. The main difference from the classical scenarios is that the artificial covariates in the homogenized models are not identically distributed as the natural covariates in the original models, consequently, the related theoretical properties are distinct from the classical ones. The asymptotical properties of the online updating statistics are established, which show that the new method can achieve estimation efficiency and oracle property, without any constraint on the number of data batches. The behavior of the method is further illustrated by various numerical examples from simulation experiments.


\end{abstract}

\baselineskip=20pt

\newpage

\setcounter{equation}{0}
\section{Introduction}

In classical regression models, the covariate set is supposed to be unchanged in the entire modeling procedure. Midway through the big data streams, however, the covariate set often changes according to, for example the change of related conditions or advances in technology. For instance, U.S. carriers were not required to report
causes of flight delays to the Bureau of Transportation Statistics until June 2003, the related variables on the causes of flight delays were available only after June 2003. Consequently, the risk analysis of airline delays needs to consider the issue of variable change midway through the data streams (e.g., Rupp, 2007). In financial analyses at the company level, new companies
become public every month, the adaptive financial analyses need to consider the ever-growing information from the new companies. For more real-world examples on the issue, see, e.g., Certo (2003), Desyllas and Sako (2013) and Wang et al. (2018).

In big data streams, data arrive in streams and blocks as  ${\cal D}_j, j=1,2,\cdots$, and in block ${\cal D}_j$ there are $n_j$ independent observations $\bm d_{ji}, i=1,\cdots,n_j$.
Consequently, the data can exceed even a super computer's memory when the number of the blocks is large enough. In this case, a crucial issue is how to address statistics in an online updating framework, without storage
requirement for previous raw data. Up to now, various statistical and computing
methodologies that enable us to sequentially update certain statistics have been proposed in the existing literature (see the references given later). Even so, however, there is rare work on formulating online updating statistics for the models with gradually changing sets of covariates in the big data streams. Under such a situation, the main difficulties in constructing online updating statistics are what follows.
\begin{itemize}\item  After the original model (or the covariate set) has changed, the estimator obtained from the previous model may be a biased estimator for the parameter in the current updated model. This indicates that previous estimators (or previous statistics) cannot be directly employed in the procedure of establishing the online updating statistics for the current updated model. This violates the general rule of online updating.
\item Because the model may be incrementally updated midway through the big data streams, and the accumulated error may increase gradually, it is difficult to formulate the incremental online updating strategy, and furthermore it is difficult to achieve the estimation consistency, efficiency and oracle property.
\end{itemize}
The goal of this paper is to solve these issues. To this end, we introduce a homogenous version to express the heterogenous updating models. We first consider the following linear regressions:
\begin{eqnarray}\label{(intro-1)}&&  y_{ji}={\bm x}_{ji}^T{\bm\beta}^0+\varepsilon_{ji}, i=1,\cdots,n_j,j=1,\cdots,k;\\ \label{(intro-1-1)} &&y_{ji}={\bm x}_{ji}^T{\bm\beta}+{\bm z}_{ji}^T{\bm\theta}+\epsilon_{ji}, i=1,\cdots,n_j,j=k+1,\cdots  \end{eqnarray}
In the above, model (\ref{(intro-1-1)}) could be regarded as an updated version of model (\ref{(intro-1)}) after the observation block ${\cal D}_k$. To eliminate the heterogeneity between the regression coefficients $\bm\beta^0$ and $\bm\beta$, we use the relationship between model (\ref{(intro-1)}) and model (\ref{(intro-1-1)}) to form the following
homogenized models:
\begin{eqnarray*} && M_1(\bm\beta,\bm\theta): y_{ji}={\bm x}_{ji}^T{\bm\beta}+{\bm x}_{ji}^T\bm B\bm\theta+\varepsilon_{ji},i=1,\cdots,n_j,j=1,\cdots,k;\\&& M_2(\bm\beta,\bm\theta): y_{ji}={\bm x}_{ji}^T{\bm\beta}+{\bm z}_{ji}^T{\bm\theta}+\epsilon_{ji}, i=1,\cdots,n_j,j=k+1,\cdots,\end{eqnarray*} where $\bm B$ is a known or estimable matrix. The models $M_1(\bm\beta,\bm\theta)$ and $M_2(\bm\beta,\bm\theta)$ are of homogeneity, in other words, the two models only contain the same parameter vectors $\bm\beta$ and $\bm\theta$. This treatment provides an opportunity to use all the compressed data to establish online updating statistics under updated model (\ref{(intro-1-1)}).

The methodology based on the technique of homogenized modeling has the following salient features:
\begin{itemize}\item [1)]
This is a unified strategy in the sense that the homogenized models $M_1(\bm\beta,\bm\theta)$ and $M_2(\bm\beta,\bm\theta)$ can be applied to establish various online updating statistics, for example, online updating parameter estimation and online updating test statistics in linear models and multiple updating models.
\item[2)] The resulting online updating estimator can achieve the estimation efficiency and oracle property. These properties are always satisfied, without any
constraint on the number of data batches.
\item[3)] Because of unified homogenized representations, the estimator is a type of least squares estimators. Thus, the procedures of constructing estimation and test statistics are simple, structurally and computationally.
\end{itemize}
The main difference from the standard regression analysis is that when $\bm B$ is replaced with its estimator $\widehat{\bm B}$, the artificial covariate ${\bm x}_{ji}^T\widehat{\bm B}\bm$ in model $M_1(\bm\beta,\bm\theta)$ are not identically distributed as the natural covariate ${\bm z}_{ji}$ in model $M_2(\bm\beta,\bm\theta)$. Thus, the theoretical properties of the resulting updating estimators are different from those in the standard regression analysis.

Before ending the introduction, we briefly summarize the existing works and issues mainly on the online updating methodologies for the models in big data streams.
In the age of big data, the explosive growth of data brings  new challenges for many classical statistical problems. Among these challenges, the major one is that data storage and analysis by standard computers are hardly feasible. Up to now, various statistical and computing methodologies have been proposed to deal with the problem.
The main strategies include sub-sampling (see, e.g., Liang et al., 2013; Kleiner et al., 2014; Maclaurin and Adams, 2014; Ma et al., 2015), divide and conquer (see, e.g., Lin and Xi, 2011; Neiswanger et al. 2013; Scott et al., 2013; Chen and Xie, 2014; Song and Liang, 2014; Pillonetto, et al., 2019), and the online updating (see, e.g., Schifano et al., 2016; Wang et al, 2018; Xue et al, 2019; Luo and Song, 2020, Cai et al., 2020).

Simply speaking, the strategy of online updating in big data streams is to address statistics in an updating framework, without storage requirement for previous raw data. In recent years, a large body of literature has emerged on studying the online updating estimation and inference in big data streams. It is known that some simple statistics, such as sample mean, least squares estimator in linear regression and N-W estimator in nonparametric regression, can be directly expressed as online updating form (see, e.g., Schifano, et al., 2016; Bucak and Gunsel, 2009; Nion and Sidiropoulos, 2009).
Under most situations, the statistics are not linear functions of data, and often have no closed form expression. In these complicated cases, stochastic gradient descent algorithm and its improved versions have been proposed to update the statistics with sequentially arriving data (see, e.g., Robbins and Monro, 1951; Bordes et al., 2009; Duchi et al., 2011; Toulis et al., 2015). For further developments in updating algorithms, such as natural gradient algorithm, the on-line Newton step, and on-line quasi-Newton algorithm, see, e.g., Amari et al. (2000),  Hazan et al. (2007), Vaits et al. (2015), Hao et al. (2016), Nocedal and Wright (1999), Liu and Nocedal (1989), Schraudolph et al. (2007) and Bordes et al. (2009).

Most exiting methodologies, however, mainly focus on the case where the variable set does not change in the entire procedure of modeling. As stated above, the variable set often changes midway through the big data streams. These existing methods cannot be directly employed for formulating the online updating estimation and inference in the models with changing variable set because the estimator obtained from the previous model is a biased estimator for the parameter in the current updated model. Recently, Wang et al. (2018) proposed a bias-correction method to deal with the problem for linear and generalized linear models with variable-addition by sequentially correcting the bias of the estimator obtained in the previous procedure. For general problems in statistical inferences, the simple and unified strategy has not been introduced, the relevant oracle properties have not been obtained, to the best of our knowledge. Thus, we desire to develop a unified strategy with a simple algorithm to implement online updating estimation and inference for general updating models.

The remainder of this paper is organized in the following way. In Section 2, for linear model, a unified framework with homogenous regression coefficients is introduced, the online updating estimators of coefficients are suggested, and the online updating expressions of residual sum of squares and $F$-statistic are presented as well.
The theoretical properties of the online updating estimator are investigated in Section 3. Simulation studies are provided in Section 4 to illustrate the new method. Proofs of the theorems are relegated to Appendix.

\setcounter{equation}{0}
\section{Methodology}

In this section, we first focus on the case of one updating linear model.

\subsection{Models}

We suppose that the underlying true model is the following linear regression: \begin{eqnarray}\label{(true)}y={\bm x}^T{\bm\beta}+{\bm z}^T\bm\theta+\epsilon, \end{eqnarray} where $\bm\beta\in R^p$ and $\bm\theta\in R^q$ are unknown true parameter vectors, and the error $\epsilon$ has conditional mean $E[\epsilon|{\bm x},{\bm z}]=0$ and finite conditional variance $\sigma^2=var[\epsilon|{\bm x},{\bm z}]$. Assume $E[{\bm x}]={\bf 0}$ and $E[{\bm z}]={\bf 0}$, without loss of generality. Here the truth of the model is particularly defined as that all the relevant covariates are included in the model, and both $\bm\beta$ and $\bm\theta$ are two nonzero vectors.

In early survey, however, only covariate vector ${\bm x}\in R^p$ can be observed.  In the procedure of survey, the data arrive in sequential blocks ${\cal D}_j, j=1,2,\cdots,k$, and in block ${\cal D}_j$ there are $n_j$ independent observations of $\bm d=(y, {\bm x})$ as $\bm d_{ji}=(y_{ji},{\bm x}_{ji}), i=1,\cdots,n_j$. Under this early situation, the working model is chosen as the following linear regression:
\begin{eqnarray}\label{(eq-1)}y_{ji}={\bm x}_{ji}^T{\bm\beta}^0+\varepsilon_{ji}, \ i=1,\cdots,n_j,j\leq k, \end{eqnarray} where $\bm\beta^0\in R^p$ is an unknown parameter vector, the errors $\varepsilon_{ji},i=1,\cdots,n_j$, are mutual independent with conditional mean $E[\varepsilon_{ji}|{\bm x}_{ji}]=0$ and finite conditional variance $\sigma^2_\varepsilon=var[\varepsilon_{ji}|{\bm x}_{ji}]$. It can be verified that $\sigma^2_\varepsilon\geq\sigma^2$ (see Remark 2.1 below).

After the $k$-th block ${\cal D}_k$, the survey environment has changed,  the remainder covariate vector ${\bm z}\in R^q$ becomes available. Under such a situation, with the complete covariate vectors $\bm x$ and $\bm z$, the model returns to be the true regression:
\begin{eqnarray}\label{(eq-2)}y_{ji}={\bm x}_{ji}^T{\bm\beta}+{\bm z}_{ji}^T{\bm\theta}+\epsilon_{ji}, \ i=1,\cdots,n_j, \ j>k, \end{eqnarray} where ${\bm z}_{ji},i=1,\cdots,n_j$ for $j>k$, are independent observations of ${\bm z}$.

For better understanding above two models, we give following explanations: Although (\ref{(eq-1)}) is chosen as a working model, it is unbiased, i.e., the errors $\varepsilon_{ji}$ have zero conditional mean $E[\varepsilon_{ji}|{\bm x}_{ji}]=0$. Actually, by the unbiasedness and the definition of true model (\ref{(eq-2)}), we have ${\bm x}^T_{ji}\bm\beta^{0}={\bm x}^T_{ji}\bm\beta+E[{\bm z}^T_{ji}|{\bm x}_{ji}]{\bm\theta}$. Consequently, we have the following lemma.

\

\noindent{\bf Lemma 2.1.} {\it If $(E[{\bm x\bm x}^T])^{-1}$ exists, $E[{\bm x}]={\bf 0}$ and $E[{\bm z}]={\bf 0}$, then, the regression coefficient $\bm\beta^0$ in model (\ref{(eq-1)}) and the regression coefficient $\bm\beta$ in model (\ref{(eq-2)}) have the following relation:
$$\bm\beta^{0}=\bm\beta+{\bm B}{\bm\theta},$$ where ${\bm B}=(E[{\bm x\bm x}^T])^{-1}E[{\bm x} {\bm z}^T]$. Particularly, if ${\bm x}$ and ${\bm z}$ are uncorrelated, then
$\bm\beta^{0}=\bm\beta.$}

The above shows that the regression coefficient $\bm\beta^0$ in model (\ref{(eq-1)}) and the regression coefficient $\bm\beta$ in model (\ref{(eq-2)}) may be completely different.

\begin{rema}
	\begin{itemize}
\item [1)] Because of the distinction between $\bm\beta^0$ and $\bm\beta$, the estimator of $\bm\beta^0$ derived from the previous model (\ref{(eq-1)}) may be a biased estimator for $\bm\beta$ in the updated model (\ref{(eq-2)}). Thus, the estimator of $\bm\beta^0$ obtained from the previous model (\ref{(eq-1)}) cannot be directly employed in the inference procedure for the true model (\ref{(eq-2)}).
\item [2)] When the model is updated, a basic task is to establish the consistent updating estimators of the parameter $\bm\beta$ and $\bm\theta$, and the test statistics for model (\ref{(eq-2)}). Note that although the two models (\ref{(eq-1)}) and (\ref{(eq-2)}) are different, they contain a common covariate $\bm x$. Thus, an important issue is to use the common ground to improve the online updating estimators of $\bm\beta$ and $\bm\theta$, and construct updating test statistics.   \end{itemize}
\end{rema}

\subsection{Homogenized models and parameter estimations}
Before introducing the homogenization technique, we first give following notations:
for all $l\leq j$, let
${\bm X}_l=({\bm x}_{l1},\cdots,{\bm x}_{ln_l})^T$, ${\bm Z}_l=({\bm z}_{l1},\cdots,{\bm z}_{ln_l})^T$ and ${\bm y}_l=(y_{l1},\cdots,y_{ln_l})^T$. Denote by $w_{1}$ and $w_{2}$ the weights that will be identified later.
For $l\leq k$, write ${\bm \varepsilon}_l=(\varepsilon_{l1},\cdots,\varepsilon_{ln_l})^T$ and
${\bm V}^{X\bm\varepsilon}_k=\sum_{l=1}^k w_{1}{\bm X}^T_l{\bm \varepsilon}_l$.
For all $j$, let $N_j=\sum_{l=1}^jn_l$ and
\begin{eqnarray}\label{(notation-1)} \nonumber&& {\bm V}^X_j=\left\{\begin{array}{ll}\sum_{l=1}^j w_{1}{\bm X}^T_l{\bm X}_l,& \mbox{if } j\leq k\vspace{1ex} \\\sum_{l=1}^k w_{1}{\bm X}^T_l{\bm X}_l+\sum_{l=k+1}^j w_{2}{\bm X}^T_l{\bm X}_l &  \mbox{if } j> k, \end{array}\right.  \vspace{1ex} \\ && {\bm V}^{Xy}_j=\left\{\begin{array}{ll}\sum_{l=1}^j w_{1}{\bm X}^T_l{\bm y}_l,& \mbox{if } j\leq k \vspace{1ex} \\\sum_{l=1}^k w_{1}{\bm X}^T_l{\bm y}_l+\sum_{l=k+1}^j w_{2}{\bm X}^T_l{\bm y}_l &  \mbox{if } j> k. \end{array}\right.\end{eqnarray}  For  $j> k$, write $M_k=\sum_{l=k+1}^jn_l$ and
\begin{eqnarray}\label{(notation-2)}&&\nonumber  {\bm \epsilon}_l=(\epsilon_{l1},\cdots,\epsilon_{ln_l})^T \mbox{ if } l> k,\\&&\nonumber
{\bm V}_{k+1,j}^{X}=\sum_{l=k+1}^j w_{2}{\bm X}^T_l{\bm X}_l,{\bm V}_{k+1,j}^{Z}=\sum_{l=k+1}^j w_{2}{\bm Z}^T_l{\bm Z}_l,
{\bm V}_{k+1,j}^{XZ}=\sum_{l=k+1}^j w_{2}{\bm X}_l^T{\bm Z}_l, \\&&\nonumber {\bm V}_{k+1,j}^{Xy}=\sum_{l=k+1}^j w_{2}{\bm X}^T_l{\bm y}_l,{\bm V}_{k+1,j}^{Zy}=\sum_{l=k+1}^j w_{2}{\bm Z}^T_l{\bm y}_l,\\&&
{\bm V}_{k+1,j}^{Z\bm\epsilon}=\sum_{l=k+1}^j w_{2}{\bm Z}^T_l{\bm\epsilon}_l,{\bm V}^{X\bm\epsilon}_{k+1,j}=\sum_{l=k+1}^j w_{2}{\bm X}^T_l{\bm \epsilon}_l.\end{eqnarray}

For the case of $j\leq k$, the weights $w_{1}$ is chosen as $ w_{1}=1$, the cumulative coefficient estimator $\widehat{\bm\theta}_j$ of ${\bm\theta}$ should be zero vector $\bm 0$, and the cumulative coefficient estimator of ${\bm\beta}$ (i.e., $\bm\beta^0$) can be chosen as the least squares estimator $\widehat{\bm\beta}^0_j$ computed on the data sets from block ${\cal D}_1$ to block ${\cal D}_j$. By these choices, the estimator $\widehat{\bm\beta}^0_j$ can be expressed as the following online updating form:
\begin{eqnarray}\label{(solution-1)}\widetilde {\bm\beta}^0_j=\left({\bm X}^T_j{\bm X}_j+{\bm V}^X_{j-1}\right)^{-1}\left({\bm X}^T_j{\bm X}_j\widehat{\bm\beta}^0_{n_j}+{\bm V}_{j-1}^X\widetilde{\bm\beta}^0_{j-1}\right),\ j\leq k,\end{eqnarray} where $\widehat{\bm\beta}^0_{n_j}=({\bm X}^T_j{\bm X}_j)^{-1}{\bm X}^T_j{\bm y}_j$, the least squares estimator of $\bm\beta^0$ computed only on the block ${\cal D}_j$, and $\widetilde{\bm\beta}_0$ and ${\bm V}^X_0$ are respectively zero vector and zero matrix.

For the case of $j>k$, we need to develop new estimation strategies. To avoid multivariate nonparametric estimation, we suppose that quadratic variable $\bm z \bm z^T$ is uncorrelated with $\bm x$, or the correlation between $\bm z \bm z^T$ and $\bm x$ can be ignored, although the correlation between $\bm x$ and $\bm z$ is taken into account in our paper. With the assumption, the error of model (\ref{(eq-1)}) has conditional variance $\sigma^2_\varepsilon=\bm\theta^T E[\bm z \bm z^T]\bm\theta+\sigma^2$ given $\bm x$. Note that the error of model (\ref{(eq-2)}) has conditional variance $\sigma^2$ given $\bm x$ and $\bm z$. Then, $w_{1}$ is chosen as $w_{1}=\overline\sigma^{-1}_\varepsilon$ and $w_{2}$ is chosen as $w_{2}={\sigma}_0^{-1}$, where $\overline\sigma^{2}_\varepsilon={\bm\theta}_0^{T} E_0[\bm z\bm z^T]{\bm\theta}_0+{\sigma}_0^{2}$, and ${\sigma}_0^{2}$, ${\bm\theta}_0$ and $E_0[\bm z\bm z^T]$ are initial choices of ${\sigma}^{2}$, ${\bm\theta}$ and $E[\bm z\bm z^T\bm]$ respectively. These initial choices may be non-random, or empirical estimators computed on ${\cal D}_{k+1}$. We consider the following two cases.

{\bf Case 1: ${\bm x}$ and ${\bm z}$ are uncorrelated.} Under such a situation, we have $\bm\beta^0=\bm\beta$, as shown in Lemma 2.1. By the relation $\bm\beta^0=\bm\beta$, and models (\ref{(eq-1)}) and  (\ref{(eq-2)}) with weights $w_{1}$ and $w_{2}$ aforementioned above, we get the following weighted models:
\begin{eqnarray}\label{(eq-4)}\nonumber &&M_1(\bm\beta,\bm\theta): \overline\sigma^{-1}_\varepsilon y_{ji}=\overline\sigma^{-1}_\varepsilon {\bm x}_{ji}^T{\bm\beta}+\overline\sigma^{-1}_\varepsilon\textbf{\emph 0}^T\bm\theta+\overline\sigma^{-1}_\varepsilon\varepsilon_{ji},i=1,\cdots,n_j, j\leq k,\\\label{(eq-5)}&& M_2(\bm\beta,\bm\theta):
{\sigma}_0^{-1} y_{ji}={\sigma}_0^{-1}{\bm x}_{ji}^T{\bm\beta}+{\sigma}_0^{-1}{\bm z}_{ji}^T{\bm\theta}+{\sigma}_0^{-1}\epsilon_{ji}, i=1,\cdots,n_j, j>k, \end{eqnarray} where $\textbf{\emph 0}$ is a zero vector.
The above models $M_1(\bm\beta,\bm\theta)$ and $M_2(\bm\beta,\bm\theta)$ are homogenized, having the same regression coefficient vectors $\bm\beta$ and $\bm\theta$. By lest squares, we obtain the cumulative estimators of ${\bm\beta}$ and $\bm\theta$ as
\begin{eqnarray}\label{(estimator-1)}
\left(\begin{array}{ll}\widehat{{\bm\beta}}_j\\ \widehat{\bm\theta}_j\end{array}\right)
=\left(\begin{array}{cccc}{\bm V}^X_j&{\bm V}^{XZ}_{k+1,j}\\ {\bm V}^{ZX}_{k+1,j}&{\bm V}^{Z}_{k+1,j}
\end{array}\right)^{-1}\left(\begin{array}{ll}{\bm V}^{Xy}_j
\\ {\bm V}^{Zy}_{k+1,j}\end{array}\right)\ \mbox{ for } j>k.\end{eqnarray}
The estimators in (\ref{(estimator-1)}) can be further expressed as the following online updating form:
\begin{eqnarray}\label{(updating-1)}
&&\hspace{-8ex}\left(\begin{array}{ll}\widetilde{{\bm\beta}}_j
\\\widetilde{\bm\theta}_j\end{array}\right)
=\left(\begin{array}{cccc}{\bm V}^X_{j-1}+{w_{2}\bm X}_j^T{\bm X}_j&{\bm V}^{XZ}_{k+1,j-1}+w_{2}{\bm X}_j^T{\bm Z}_j\\ {\bm V}^{ZX}_{k+1,j-1}+{w_{2}\bm Z}_j^T{\bm X}_j&{\bm V}^{Z}_{k+1,j-1}+w_{2}{\bm Z}_j^T{\bm Z}_j
\end{array}\right)^{-1}\\&&\hspace{-4ex}\nonumber\times\left(\left(\begin{array}{cccc}w_{2} {\bm X}_j^T{\bm X}_j& w_{2}{\bm X}_j^T{\bm Z}_j\\ w_{2}{\bm Z}_j^T{\bm X}_j&w_{2} {\bm Z}_j^T{\bm Z}_j
\end{array}\right)\left(\begin{array}{ll}\widehat{\bm\beta}_{n_j}\\ \widehat{\bm\theta}_{n_j}\end{array}\right)+\left(\begin{array}{cccc}{\bm V}^X_{j-1}&{\bm V}^{XZ}_{k+1,j-1} \\ {\bm V}^{ZX}_{k+1,j-1}&{\bm V}^{Z}_{k+1,j-1}
\end{array}\right)\left(\begin{array}{ll}\widetilde{{\bm\beta}}_{j-1}\\
\widetilde{\bm\theta}_{j-1}\end{array}\right)\right)\end{eqnarray} for $j>k$, where $$\left(\begin{array}{ll}\widehat{\bm\beta}_{n_j}\\ \widehat{\bm\theta}_{n_j}\end{array}\right)=\left(\begin{array}{cccc} {w_{2}\bm X}_j^T{\bm X}_j&w_{2} {\bm X}_j^T{\bm Z}_j\\ w_{2}{\bm Z}_j^T{\bm X}_j& w_{2}{\bm Z}_j^T{\bm Z}_j
\end{array}\right)^{-1}\left(\begin{array}{ll}{w_{2}\bm X}_j^T{\bm y}_j\\ w_{2}{\bm Z}_j^T{\bm y}_j\end{array}\right),$$
the least squares estimator computed on the $j$-th block ${\cal D}_j$.

\vspace{1ex}

\begin{rema}
	\begin{itemize}\item[1)]
The estimators in (\ref{(updating-1)}) contain the information from the previous data sets ${\cal D}_l$ with $l\leq k$. Thus, it can be expected that the estimators can be improved. By the theorems in the next section, we have that the online updating estimators $\widetilde{\bm\beta}_j$ and $\widetilde{\bm\theta}_j$ achieve the oracle convergence rates of orders $O_p(1/\sqrt{N_j})$ and  $O_p(1/\sqrt{M_k})$ respectively, and $\widetilde{\bm\theta}_j$ behaves as the same as the online updating form $\widetilde{\bm\theta}^{naive}_j$ of the naive estimator $\widehat{\bm\theta}^{naive}_j$ obtained by minimizing $$\sum_{l=k+1}^j\sum_{i=1}^{n_l}(y_{li}-{\bm x}_{li}^T{\bm\beta}-{\bm z}_{li}^T{\bm\theta})^2$$ for $\bm\theta$. Therefore, for simplicity, the final online updating estimators of $\bm\beta$ and $\bm\theta$ can be chosen as $\widetilde{\bm\beta}_j$ and $\widetilde{\bm\theta}^{naive}_j$ respectively.
\item[2)]
The statistics in (\ref{(updating-1)}) are of online updating form because they only involve the current data $w_{2}{\bm X}_j^T{\bm X}_j$, ${w_{2}\bm X}^T_j{\bm Z}_j$ and $w_{2}{\bm Z}_j^T{\bm Z}_j$, and the current estimators $\widehat{\bm\beta}_{n_j}$ and $\widehat{\bm\theta}_{n_j}$, together with the quantities ${\bm V}^X_{j-1}$, ${\bm V}^{XZ}_{k+1,j-1}$, ${\bm V}^Z_{k+1,j-1}$, $ {\bm V}^{Zy}_{k+1,j-1}$ and the estimators $\widetilde{{\bm\beta}}_{j-1}$ and $\widetilde{\bm\theta}_{j-1}$ from the previous accumulation.
\end{itemize}
\end{rema}

{\bf Case 2: ${\bm x}$ and ${\bm z}$ are correlated.} In this case, the the homogenization method proposed in Case 1 does not work because it ignores the correlation between the data ${\bm x}_{li}$ and ${\bm z}_{li}$ and the correlation may result in that the online updating estimators $\widetilde{{\bm\beta}}_j$ and $\widetilde{{\bm\theta}}_j$ in (\ref{(updating-1)}) have a non-negligible bias (see Remark 3.1 given in the next section).

According to the error variances of model (\ref{(eq-1)}) and (\ref{(eq-2)}), we set the weights as $w_{1}=\overline\sigma^{-1}_\varepsilon$ and $w_{2}={\sigma}_0^{-1}$ as in Case 1.
By Lemma 2.1, we replace $\bm\beta^{0}$ with $\bm\beta+{\bm B}{\bm\theta}$. We then get the following weighted models:
\begin{eqnarray}\label{(correct-eq-4)}\nonumber &&M_1(\bm\beta,\bm\theta): \overline\sigma^{-1}_\varepsilon y_{ji}=\overline\sigma^{-1}_\varepsilon {\bm x}_{ji}^T{\bm\beta}+\overline\sigma^{-1}_\varepsilon{\bm x}_{ji}^T\widehat{\bm B}\bm\theta+\overline\sigma^{-1}_\varepsilon\varepsilon_{ji},i=1,\cdots,n_j, j\leq k,\\\label{(eq-5)}&& M_2(\bm\beta,\bm\theta):
{\sigma}_0^{-1} y_{ji}={\sigma}_0^{-1}{\bm x}_{ji}^T{\bm\beta}+{\sigma}_0^{-1}{\bm z}_{ji}^T{\bm\theta}+{\sigma}_0^{-1}\epsilon_{ji}, i=1,\cdots,n_j, j>k, \end{eqnarray}
where $\widehat{\bm B}$ is an empirical estimator of ${\bm B}$ computed on ${\cal D}_{k+1}$. The above models $M_1(\bm\beta,\bm\theta)$ and $M_2(\bm\beta,\bm\theta)$ are of homogeneity, having the same regression coefficient vectors $\bm\beta$ and $\bm\theta$.
Then, the cumulative estimators can be expressed as
\begin{eqnarray}\label{(correct-estimator-1)}
\left(\begin{array}{ll}\widehat{{\bm\beta}}_j\\ \widehat{\bm\theta}_j\end{array}\right)
=\left(\begin{array}{cccc}{\bm V}^X_j&{\bm V}^{X}_{k}\widehat{\bm B}+{\bm V}^{XZ}_{k+1,j}\\ {\bm V}^{ZX}_{k+1,j}&{\bm V}^{Z}_{k+1,j}
\end{array}\right)^{-1}\left(\begin{array}{ll}{\bm V}^{Xy}_j
\\ {\bm V}^{Zy}_{k+1,j}\end{array}\right)\ \mbox{ for } j>k.\end{eqnarray}
Similar to (\ref{(updating-1)}), the above estimators can be expressed as the following online updating form:
\begin{eqnarray}\label{(correct-updating-1)}
&&\hspace{-8ex}\left(\begin{array}{ll}\widetilde{{\bm\beta}}_j
\\\widetilde{\bm\theta}_j\end{array}\right)
=\left(\begin{array}{llll}{\bm V}^X_{j-1}+w_{2}{\bm X}_j^T{\bm X}_j&{\bm V}^{X}_{k}\widehat{\bm B}+{\bm V}^{XZ}_{k+1,j-1}+w_{2}{\bm X}_j^T{\bm Z}_j\\ {\bm V}^{ZX}_{k+1,j-1}+w_{2}{\bm Z}_j^T{\bm X}_j&{\bm V}^{Z}_{k+1,j-1}+w_{2}{\bm Z}_j^T{\bm Z}_j
\end{array}\right)^{-1}\\&&\hspace{-4ex}\nonumber\times\left(\left(\begin{array}{llll}w_{2} {\bm X}_j^T{\bm X}_j& w_{2}{\bm X}_j^T{\bm Z}_j\\ w_{2}{\bm Z}_j^T{\bm X}_j&w_{2} {\bm Z}_j^T{\bm Z}_j
\end{array}\right)\left(\begin{array}{ll}\widehat{\bm\beta}_{n_j}\\ \widehat{\bm\theta}_{n_j}\end{array}\right)+\left(\begin{array}{llll}{\bm V}^X_{j-1}&{\bm V}^{XZ}_{k+1,j-1} \\ {\bm V}^{ZX}_{k+1,j-1}&{\bm V}^{Z}_{k+1,j-1}
\end{array}\right)\left(\begin{array}{ll}\widetilde{{\bm\beta}}_{j-1}\\
\widetilde{\bm\theta}_{j-1}\end{array}\right)\right)\end{eqnarray} for $j>k$, where $$\left(\begin{array}{ll}\widehat{\bm\beta}_{n_j}\\ \widehat{\bm\theta}_{n_j}\end{array}\right)=\left(\begin{array}{cccc} w_{2}{\bm X}_j^T{\bm X}_j& w_{2}{\bm X}_j^T{\bm Z}_j\\ {w_{2}\bm Z}_j^T{\bm X}_j& w_{2}{\bm Z}_j^T{\bm Z}_j
\end{array}\right)^{-1}\left(\begin{array}{ll}w_{2}{\bm X}_j^T{\bm y}_j\\w_{2} {\bm Z}_j^T{\bm y}_j\end{array}\right).$$

Actually the estimators (\ref{(estimator-1)}) and its online updating form (\ref{(correct-estimator-1)}) are also applied to Case 1 because in this case the estimator $\widehat{\bm B}=\bm 0$ in some sense, consequently, the estimators (\ref{(estimator-1)}) and (\ref{(correct-estimator-1)}) are equal to (\ref{(estimator-1)}) and (\ref{(updating-1)}) in the same sense.

\begin{rema}
	By the same argument as used in Remark 2.2, for simplicity, the final online updating estimators of $\bm\beta$ and $\bm\theta$ can be respectively chosen as $\widetilde{\bm\beta}_j$ and $\widetilde{\bm\theta}^{naive}_j$ under Case 2.
\end{rema}

\subsection{Online updating forms of other statistics}

\subsubsection{Online updating forms of residual sum of squares}

The residual sum of squares is a key statistic for statistical inference. We first
discuss the online updating form of residual sum of squares. According to the homogenized models in (\ref{(eq-4)}), under Case 1, the weighted covariates can be written uniformly by
\begin{eqnarray*}{\bm s}_{li}=\left\{\begin{array}{ll}\left(\begin{array}{cc}\overline\sigma^{-1}_\varepsilon{\bm x}_{li}\\ \textbf{\emph 0}\end{array}\right),& \mbox{if } l\leq N_k\vspace{1ex} \\\left(\begin{array}{cc}\sigma_0^{-1}{\bm x}_{li}\\ \sigma_0^{-1}{\bm z}_{li}\end{array}\right), &  \mbox{if } l> N_k. \end{array}\right. \mbox{ for } j>k.\end{eqnarray*} Under Case 2, according to the homogenized models in (\ref{(correct-eq-4)}), the weighted covariates can be uniformly expressed by
\begin{eqnarray*}{\bm s}_{li}=\left\{\begin{array}{ll}\left(\begin{array}{cc}\overline\sigma^{-1}_\varepsilon{\bm x}_{li}\\
\overline\sigma^{-1}_\varepsilon{\bm x}_{li}^T\widehat{\bm B}\end{array}\right),& \mbox{if } l\leq N_k\vspace{1ex} \\\left(\begin{array}{cc}\sigma_0^{-1}{\bm x}_{li}\\ \sigma_0^{-1}{\bm z}_{li}\end{array}\right), &  \mbox{if } l> N_k, \end{array}\right. \mbox{ for } j>k.\end{eqnarray*}
Let $\widetilde{\bm \eta}_j=\left(\begin{array}{ll}\widetilde{{\bm\beta}}_j\\ \widetilde{{\bm\theta}}_j\end{array}\right)$, $\widehat{\bm \eta}_j=\left(\begin{array}{ll}\widehat{\bm\beta}_{j}\\ \widehat{\bm\theta}_{j}\end{array}\right)$ and $\widehat{\bm \eta}_{n_j}=\left(\begin{array}{ll}\widehat{\bm\beta}_{n_j}\\ \widehat{\bm\theta}_{n_j}\end{array}\right)$.
Then, the residual sum of squares is defined by
\begin{eqnarray}\label{(residual)}\nonumber SSE_j&=&\sum_{l=1}^{N_k}\overline\sigma^{-2}_\varepsilon{\bm y}_l^T{\bm y}_l+\sum_{l=N_k+1}^{j}\sigma_0^{-2}{\bm y}_l^T{\bm y}_l\\&&-\left(\sum_{l=1}^j{\bm S}_l^T{\bm S}_l\widehat{\bm \eta}_{n_l}\right)^T\left(\sum_{l=1}^j{\bm S}_l^T{\bm S}_l\right)^{-1}\left(\sum_{l=1}^j{\bm S}_l^T{\bm S}_l\widehat{\bm \eta}_{n_l}\right),\end{eqnarray} where ${\bm S}_l=({\bm s}_{l1},\cdots,{\bm s}_{ln_l})^T$.
By the technique of Schifano, et al. (2016), the above can be recast as the online updating form:
\begin{eqnarray}\label{(residual-1)}SSE_j=SSE_{j-1}+SSE_{n_j}+\widetilde{\bm \eta}^T_{j-1}{\bm V}_{j-1}\widetilde{\bm \eta}_{j-1}+\widehat{\bm \eta}^T_{n_j}{\bf S}^T_{j}{\bf S}_j\widehat{\bm \eta}_{n_j}-\widetilde{\bm \eta}^T_{j}{\bm V}_{j}\widetilde{\bm \eta}_{j},\end{eqnarray} where $SSE_{n_j}$ is the residual sum of squares from ${\cal D}_j$.
The representation in (\ref{(residual-1)}) is of online updating form because it only involves the current data ${\bm S}_j^T{\bf S}_j$ and current residual sum of squares $SSE_{n_j}$, together with the quantities ${\bm V}_{j-1}$, $SSE_{j-1}$ and the estimator $\widetilde{\bm \eta}_{j-1}$ from the previous accumulation.

\subsubsection{Online updating forms of $F$-statistic}

Actually, most of important statistics such as $F$-statistic and $t$-statistic also can be expressed as online updating form via the homogenized models (\ref{(eq-4)}) and (\ref{(correct-eq-4)}) together with the corresponding estimators (\ref{(updating-1)}) and (\ref{(correct-updating-1)}).

For example, it is crucial to test if model (\ref{(eq-2)}) is true when the new covariate vector $\bm z$ becomes available after the $k$-th block ${\cal D}_k$. This issue is equivalent to testing
\begin{eqnarray}\label{(test-1)}H:\bm\theta=\emph{\bf 0}.\end{eqnarray} For Case 1 where $\bm x$ and $\bm z$ are uncorrelated, by the corresponding homogenized models in (\ref{(eq-4)}), the $F$-statistic should be
\begin{eqnarray}\label{(test-2)}F_j=\frac{\widetilde{\bm\theta}_j^T\left({\bm V}^{Z}_{k+1,j}-{\bm V}^{ZX}_{k+1,j}\left({\bm V}^{X}_{j}\right)^{-1}{\bm V}^{XZ}_{k+1,j}\right)\widetilde{\bm\theta}_j/q}{SSE_j/(N_j-q)} \ \mbox{ for } j>k.\end{eqnarray} Note that the factor ${\bm V}^{Z}_{k+1,j}-{\bm V}^{ZX}_{k+1,j}({\bm V}^{X}_{j})^{-1}{\bm V}^{XZ}_{k+1,j}$ in the numerator has the following online updating form:
\begin{eqnarray}\label{(online-test)}&&\hspace{-3ex}{\bm V}^{Z}_{k+1,j-1}+{w_{2}\bm Z}_j^T{\bm Z}_j\\&&\hspace{-3ex}\nonumber -({\bm V}^{ZX}_{k+1,j-1}+w_{2}{\bm Z}_j^T{\bm X}_j)\left({\bm V}^X_{j-1}+w_{2}{\bm X}_j^T{\bm X}_j\right)^{-1}({\bm V}^{XZ}_{k+1,j-1}+w_{2}{\bm X}_j^T{\bm Z}_j).\end{eqnarray} Based on these online updating forms in (\ref{(updating-1)}), (\ref{(residual-1)}) and (\ref{(online-test)}), the $F$-statistic $F_j$ in (\ref{(test-2)}) is of online updating form.

For Case 2, by the corresponding homogenized models in (\ref{(correct-eq-4)}), the $F$-statistic is defined by
\begin{eqnarray}\label{(test-3)} F^c_j=\frac{\widetilde{\bm\theta}_j^T\left({\bm V}^{Z}_{k+1,j}-{\bm V}^{ZX}_{k+1,j}\left({\bm V}^{X}_{j}\right)^{-1}({\bm V}^{X}_{k}\widehat{\bm B}+{\bm V}^{XZ}_{k+1,j})\right)\widetilde{\bm\theta}_j/q}{SSE_j/(N_j-q)} \ \mbox{ for } j>k.\end{eqnarray}
By the same argument used above, the $F$-statistic $F^c_j$ in (\ref{(test-3)}) can be expressed as the online updating form as well. Furthermore, the $F$-statistic $F^c_j$ in (\ref{(test-3)}) is also applied to Case 1 because in this case $\widehat{\bm B}=\bm 0$ in some sense, implying $F^c_j=F_j$.

When $\epsilon\sim N(0,\sigma^2)$, and ${\sigma}_0^{2}$, ${\bm\theta}_0$ and $E_0[\bm z\bm z^T\bm]$ are chosen to be consistent estimators, it can be guaranteed by Theorem 3.4 given below that under $H$, the test statistics have the following asymptotic distributions:
\begin{eqnarray}\label{(test-4)}&&F_j\stackrel{d}\longrightarrow F_{q,N_j-q} \ \mbox{in Case 1,} \\\label{(test-4-1)} && F^c_j\stackrel{d}\longrightarrow F_{q,N_j-q}\ \mbox{ in both Case 1 and Case 2},\end{eqnarray} where the notation ``$\stackrel{d}\longrightarrow$" stands for the convergence in distribution, and $F_{q,N_j-q}$ denotes the $F$-distribution on $q$ and $N_j-q$ degrees of freedom. Consequently, under Case 1, for given $0<\alpha<1$, if
$$F_j>F_{q,N_j-q}(1-\alpha),$$ we reject $H$, implying that model (\ref{(eq-2)}) is true. Similarly, under both Case 1 and Case 2, if
$$F^c_j>F_{q,N_j-q}(1-\alpha),$$ we reject $H$, implying that model (\ref{(eq-2)}) is true as well.

\subsection{Extension}

To extend the method to multiple updating case, we here mainly consider the following twice updating linear models: \begin{eqnarray}\label{(extension-1)}&&  y_{ji}={\bm x}_{ji}^T{\bm\beta}^0+\varepsilon_{ji}, i=1,\cdots,n_j,j=1,\cdots,k;\\ \label{(extension-2)} &&y_{ji}={\bm x}_{ji}^T{\bm\beta}^1+{\bm z}_{ji}^T{\bm\theta}^1+\epsilon^1_{ji}, i=1,\cdots,n_j,j=k+1,\cdots,k+m; \\ \label{(extension-3)} &&y_{ji}={\bm x}_{ji}^T{\bm\beta}^2+{\bm z}_{ji}^T{\bm\theta}^2+{\bm w}_{ji}^T{\bm\gamma}^2+\epsilon^2_{ji}, i=1,\cdots,n_j,j=k+m+1,\cdots, \end{eqnarray} where the errors have conditional expectation zero respectively given ${\bm x}_{ji}$, $({\bm x}_{ji}^T,{\bm z}_{ji}^T)^T$, and $({\bm x}_{ji}^T,{\bm z}_{ji}^T,{\bm w}_{ji}^T)^T$. In the above, model (\ref{(extension-3)}) is regarded as the true model that contains complete covariates ${\bm x}_{ji}$, ${\bm z}_{ji}$ and ${\bm w}_{ji}$, and is the twice updating version of the initial model (\ref{(extension-1)}). By the same argument as used in Lemma 2.1, we have
\begin{eqnarray}\nonumber&&\bm\beta^{0}=\bm\beta^2+{\bm B}{\bm\theta}^2+{\bm C}{\bm\gamma}^2\ \mbox{ and } \ \left(\begin{array}{cc}\bm\beta^1\\\bm\theta^1\end{array}\right)=
\left(\begin{array}{cc}\bm\beta^2\\\bm\theta^2\end{array}\right)+\bm D\bm\gamma^2,\end{eqnarray} where ${\bm B}=(E[{\bm x\bm x}^T])^{-1}E[{\bm x} {\bm z}^T]$, ${\bm C}=(E[{\bm x\bm x}^T])^{-1}E[{\bm x} {\bm w}^T]$ and $${\bm D}=\left(E\left[\left(\begin{array}{cc}{\bm x}\\{\bm z}\end{array}\right)\left(\begin{array}{cc}{\bm x}\\{\bm z}\end{array}\right)^T\right]\right)^{-1}E\left[\left(\begin{array}{cc}{\bm x}\\{\bm z}\end{array}\right) {\bm w}^T\right].$$ Then, after twice update, the heterogenous models (\ref{(extension-1)}), (\ref{(extension-2)}) and (\ref{(extension-3)}) can be expressed respectively as the following homogenized forms
\begin{eqnarray}\label{(extension-1-1)} &&\hspace{-7ex} y_{ji}= {\bm x}_{ji}^T{\bm\beta}^2+ {\bm x}_{ji}^T\widehat{\bm B}\bm\theta^2+{\bm x}_{ji}^T\widehat{\bm C}\bm\gamma^2+ \varepsilon_{ji},i=1,\cdots,n_j, 1 \leq j \leq k;\\\label{(extension-2-1)}&&\hspace{-7ex}
y_{ji}={\bm x}_{ji}^T{\bm\beta}^2+{\bm z}_{ji}^T{\bm\theta}^2+\left(\begin{array}{c}{\bm x}\\{\bm z}\end{array}\right)^T{\widehat{\bm D}}{\bm\gamma}^2+\epsilon^1_{ji}, i=1,\cdots,n_j,  k+1\leq j\leq k+m;
\\ \label{(extension-3-1)} &&\hspace{-7ex}y_{ji}={\bm x}_{ji}^T{\bm\beta}^2+{\bm z}_{ji}^T{\bm\theta}^2+{\bm w}_{ji}^T{\bm\gamma}^2+\epsilon^2_{ji}, i=1,\cdots,n_j,j\geq k+m+1,\end{eqnarray} where $\widehat{\bm B}$, $\widehat{\bm C}$ and $\widehat{\bm D}$ are the empirical estimators of $\bm B$, $\bm C$ and $\bm D$ respectively.
The three models above are of homogeneity, containing the same parameters ${\bm\beta}^2$, ${\bm\theta}^2$ and ${\bm\gamma}^2$. But the errors are of heteroscedasticity. With the weights as in
(\ref{(correct-eq-4)}), the errors can be changed to be homoscedastic. Finally, by the weighted models of (\ref{(extension-1-1)}), (\ref{(extension-2-1)}) and (\ref{(extension-3-1)}), and OLS, we can get the estimators of ${\bm\beta}^2$, ${\bm\theta}^2$ and ${\bm\gamma}^2$, and their online updating expressions, which are similar to those in
(\ref{(correct-estimator-1)}) and (\ref{(correct-updating-1)}), respectively. The details are omitted here.

Based on the rule above, the method can be extended into multiple updating case. Moreover, in principle, the method can be also extended into GLMs, because the solution to a GLM can be expressed as a solution of a weighted least squares to the corresponding heteroscedastic linear model (see, e.g., McCulloch and Searle, 2001). This issue is beyond the scope of this article, it is worth further study in the future.

\vspace{1ex}

\setcounter{equation}{0}
\section{Theoretical properties}

Although models $M_1(\bm\beta,\bm\theta)$ and $M_2(\bm\beta,\bm\theta)$ contains the same regression coefficients, the artificial covariate ${\bm x}_{ji}^T\widehat{\bm B}$ in model $M_1(\bm\beta,\bm\theta)$ are not identically distributed as the
natural covariate ${\bm z}_{ji}$ in model $M_2(\bm\beta,\bm\theta)$. Thus, the theoretical properties of the resulting updating estimators are different from those in the standard regression analysis.
We first consider the case when ${\sigma}_0^{2}$, ${\bm\theta}_0$ and $E_0[\bm z\bm z^T\bm]$ are chosen to be non-random.  We need the condition of non-randomness for achieving the standard convergence rate of order $O_p(1/\sqrt{N_j})$ (see Remark 3.3 below).
We first establish the asymptotic normality for the online updating estimators in (\ref{(updating-1)}) for model (\ref{(eq-2)}), and show the estimation efficiency, the oracle property and the adaptability to perpetual streaming data sets.

\vspace{1ex}

\begin{theorem}\label{th31}
If $\bm x$ and $\bm z$ are uncorrelated, $E[{\bm x}]={\bf 0}$, $E[{\bm z}]={\bf 0}$, matrices $E[{\bm x\bm x}^T]$ and $E[{\bm z\bm z}^T]$ are positive definite, $\frac{M_k}{N_j}\rightarrow\rho\neq 0$, and  ${\sigma}_0^{2}$, ${\bm\theta}_0$ and $E_0[\bm z\bm z^T\bm]$ are chosen to be non-random, then, the online updating estimators $\widetilde{{\bm\beta}}_j$ and $\widetilde{{\bm\theta}}_j$ in (\ref{(updating-1)}) satisfy that as $N_j\rightarrow\infty$,
\begin{eqnarray*}
\sqrt{N_j}\left(\left(\begin{array}{ll}\widetilde{{\bm\beta}}_j\\\widetilde{\bm\theta}_j\end{array}\right)
-\left(\begin{array}{ll}{\bm\beta}\\\bm\theta \end{array}\right)\right)
\stackrel{d}\longrightarrow
N\left(0, \Omega_\rho^{-1}\Phi_\rho \Omega_\rho^{-1}\right).\end{eqnarray*}
where the notation ``$\stackrel{d}\longrightarrow$" stands for convergence in distribution, and
\begin{eqnarray*}\Omega_\rho&=&\left(\begin{array}{cccc}\frac{1-\rho}{\overline\sigma_\varepsilon}E[{\bm x\bm x}^T]+\frac{\rho}{\sigma_0}E[{\bm x\bm x}^T]&{\bf 0}\\ {\bf 0}&\frac{\rho}{\sigma_0} E[{\bm z\bm z}^T]
\end{array}\right),\\ \Phi_\rho&=&\left(\begin{array}{cccc}\frac{\sigma^2_\varepsilon(1-\rho)}{\overline\sigma^2_\varepsilon}  E[{\bm x\bm x}^T]+\frac{\sigma^2\rho}{\sigma^2_0}  E[{\bm x\bm x}^T]&{\bf 0}\\ {\bf 0} &\frac{\sigma^2\rho}{\sigma_0^2} E[{\bm z\bm z}^T]
\end{array}\right).\end{eqnarray*}
\end{theorem}
The proof of the theorem is presented in Appendix. For the theorem, we have the following explanations

\begin{rema}
	\begin{itemize}
\item[1)] The theorem ensures that for the case where $\bm x$ and $\bm z$ are uncorrelated and $\frac{M_k}{N_j}\rightarrow\rho\neq 0$, the online updating estimators $\widetilde{{\bm\beta}}_j$ and $\widetilde{{\bm\theta}}_j$ in (\ref{(updating-1)}) are always $\sqrt{N_j}$-consistent and normally distributed asymptotically. These properties are always satisfied, without any constraint on the number of data batches.
\item[2)] The proof of the theorem indicates that the two estimators in (\ref{(updating-1)}) have a bias as
    $$\left(\begin{array}{cccc}\frac{1}{N_j}{\bm V}^X_j&\frac{1}{N_j}{\bm V}^{XZ}_{k+1,j}\\ \frac{1}{N_j}{\bm V}^{ZX}_{k+1,j}&\frac{1}{N_j}{\bm V}^{Z}_{k+1,j}
\end{array}\right)^{-1}\left(\begin{array}{ll}\frac{1}{N_j}{\bm V}^{XZ}_k\bm\theta
\\  {\bf 0}\end{array}\right).$$ It is obvious that this bias tends to zero as $N_j\rightarrow\infty$ provided that $\bm x$ and $\bm z$ are uncorrelated. Without these conditions, however, the bias is non-negligible.
\end{itemize}
\end{rema}

Theorem \ref{th31} needs the condition $\rho\neq 0$. Without the constraint, we have the following corollary.
\begin{coro}\label{core31}
	If $\bm x$ and $\bm z$ are uncorrelated, $E[{\bm x}]={\bf 0}$, $E[{\bm z}]={\bf 0}$, matrices $E[{\bm x\bm x}^T]$ and $E[{\bm z\bm z}^T]$ are positive definite, $\frac{M_k}{N_j}\rightarrow\rho$, and  ${\sigma}_0^{2}$, ${\bm\theta}_0$ and $E_0[\bm z\bm z^T\bm]$ are chosen to be non-random, then, the online updating estimators $\widetilde{{\bm\beta}}_j$ and $\widetilde{{\bm\theta}}_j$ in (\ref{(updating-1)}) are uncorrelated asymptotically, and satisfy that as $N_j\rightarrow\infty$,
\begin{eqnarray*}\sqrt{N_j}(\widetilde{{\bm\beta}}_j-\bm\beta)\stackrel{d}\longrightarrow
N\left(0, \left(\Omega_\rho^{(11)}\right)^{-1}\Phi^{(11)}_\rho \left(\Omega_\rho^{(11)}\right)^{-1}\right)\end{eqnarray*}
 and
\begin{eqnarray*}
\sqrt{M_k}(\widetilde{{\bm\theta}}_j-\bm\theta)\stackrel{d}\longrightarrow
N\left(0,\sigma^2  E^{-1}[{\bm z\bm z}^T]\right),\end{eqnarray*}
where \begin{eqnarray*}&&\Omega_\rho^{(11)}=\frac{1-\rho}{\overline\sigma_\varepsilon}E[{\bm x\bm x}^T]+\frac{\rho}{\sigma_0}E[{\bm x\bm x}^T],\\&& \Phi^{(11)}_\rho=\frac{\sigma^2_\varepsilon(1-\rho)}{\overline\sigma^2_\varepsilon}  E[{\bm x\bm x}^T]+\frac{\sigma^2\rho}{\sigma^2_0}  E[{\bm x\bm x}^T].\end{eqnarray*}
\end{coro}

For better understanding the theorem and corollary, we have the following remark.
\begin{rema}
\begin{itemize}
\item[1)] Theorem \ref{th31} ensures that for any $0\leq\rho \leq 1$, the online updating estimator $\widetilde{\bm\beta}_j$ achieves the oracle convergence rate of order $O_p(1/\sqrt{N_j})$ as if the estimator was obtained simultaneously using all data sets ${\cal D}_l,l=1,\cdots,j$. The online updating estimator $\widetilde{\bm\theta}_j$ has the convergence rate of order $O_p(1/\sqrt{M_k})$, and the asymptotic covariance $\sigma^2  E^{-1}[{\bm z\bm z}^T]$, which are the same as those of the naive estimator formed by the data after block ${\cal D}_k$. Thus, the improvement of the new method lies mainly in enhancing the convergence rate of the estimator $\widetilde{\bm\beta}_j$.
\item[2)] It can be seen from the proof of the theorem that we need the condition that
 ${\sigma}_0^{2}$, ${\bm\theta}_0$ and $E_0[\bm z\bm z^T\bm]$ are chosen to be non-random for achieving the standard convergence rates of order $O_p(1/\sqrt{N_j})$ and $O_p(1/\sqrt{M_k})$, respectively.
\item[3)] If $\rho=1$, the two estimators have the oracle covariances $\sigma^2_\varepsilon  E^{-1}[{\bm x\bm x}^T]$ and $\sigma^2  E^{-1}[{\bm z\bm z}^T]$, respectively. When $\rho=0$, i.e., the data come mainly from working model (\ref{(eq-1)}), asymptotically, the estimator $\widetilde{\bm\beta}_j$ is an efficient estimator of the true parameter $\bm\beta$ in true model (\ref{(eq-2)}), instead of the parameter $\bm\beta^0$ in working model (\ref{(eq-1)}).
 \item[4)] With the non-random choices, however, the estimation efficiency cannot be obtained usually. More precisely,
the choices of ${\sigma}_0^{2}$, ${\bm\theta}_0$ and $E_0[\bm z\bm z^T\bm]$ have significant influence on the estimation efficiency and particularly, when ${\sigma}_0^{2}={\sigma}^{2}$, ${\bm\theta}_0={\bm\theta}$ and $E_0[\bm z\bm z^T\bm]=E[\bm z\bm z^T\bm]$, the estimators $\widetilde{{\bm\beta}}_j$ and $\widetilde{{\bm\theta}}_j$ achieve estimation efficiency.
 \end{itemize}
\end{rema}

For Case 2, we have the following theorem.
\begin{theorem}\label{th32}
If $\bm x$ and $\bm z$ are correlated, $E[{\bm x}]={\bf 0}$, $E[{\bm z}]={\bf 0}$, matrices $E[{\bm x\bm x}^T]$ and $E[{\bm z\bm z}^T]$ are positive definite, $\frac{M_k}{N_j}\rightarrow\rho$, and  ${\sigma}_0^{2}$, ${\bm\theta}_0$ and $E_0[\bm z\bm z^T\bm]$ are chosen to be non-random, then, the online updating estimators $\widetilde{{\bm\beta}}_j$ and $\widetilde{{\bm\theta}}_j$ in (\ref{(correct-updating-1)}) satisfy that as $N_j\rightarrow\infty$,
\begin{eqnarray*}
\sqrt{N_j}\left(\left(\begin{array}{ll}\widetilde{{\bm\beta}}_j\\\widetilde{\bm\theta}_j\end{array}\right)
-\left(\begin{array}{ll}{\bm\beta}\\\bm\theta \end{array}\right)\right)
\stackrel{d}\longrightarrow
N\left(0, (\Omega^c_\rho)^{-1}\Phi^c_\rho (\Omega^c_\rho)^{-1}\right),\end{eqnarray*}
where
\begin{eqnarray*}&&\Omega^c_\rho=\left(\begin{array}{cccc}\frac{1-\rho}{\overline\sigma_\varepsilon}E[{\bm x\bm x}^T]+\frac{\rho}{\sigma_0} E[{\bm x\bm x}^T]&\frac{1-\rho}{\overline\sigma_\varepsilon}E[{\bm z\bm x}^T]+\frac{\rho}{\sigma_0} E[{\bm z\bm x}^T]\vspace{1ex}\\ \frac{\rho}{\sigma_0} E[{\bm x\bm z}^T]&\frac{\rho}{\sigma_0} E[{\bm z\bm z}^T]
\end{array}\right), \\&& \Phi^c_\rho=\left(\begin{array}{cccc}\frac{\sigma_\varepsilon^2(1-\rho)}{\overline\sigma^2_\varepsilon}  E[{\bm x\bm x}^T]+\frac{\sigma^2\rho}{\sigma^2_0} E[{\bm x\bm x}^T]&\frac{\sigma^2\rho}{\sigma^2_0} E[{\bm x\bm z}^T]\vspace{1ex}\\ \frac{\sigma^2\rho}{\sigma_0^2} E[{\bm z\bm x}^T] &\frac{\sigma^2\rho}{\sigma_0^2} E[{\bm z\bm z}^T]
\end{array}\right).\end{eqnarray*}
\end{theorem}

To better explain this theorem, we make the following remarks.
\begin{rema}
\begin{itemize}
\item[1)] It can be seen that the online updating estimators $\widetilde{\bm\beta}_j$ and $\widetilde{\bm\theta}_j$ achieve the oracle convergence rates of orders $O_p(1/\sqrt{N_j})$ and  $O_p(1/\sqrt{M_k})$, respectively.
\item[2)] The non-random choices
of ${\sigma}_0^{2}$, ${\bm\theta}_0$ and $E_0[\bm z\bm z^T\bm]$ have significant influence on the estimation efficiency and particularly, when ${\sigma}_0^{2}={\sigma}^{2}$, ${\bm\theta}_0={\bm\theta}$ and $E_0[\bm z\bm z^T\bm]=E[\bm z\bm z^T\bm]$, the estimators $\widetilde{{\bm\beta}}_j$ and $\widetilde{{\bm\theta}}_j$ in (\ref{(correct-updating-1)}) achieve estimation efficiency.\end{itemize}
\end{rema}

In order to establish the asymptotic distributions of the $F$-statistics $F_j$ and $F_j^c$, we need the condition: ${\sigma}_0^{2}$, ${\bm\theta}_0$ and $E_0[\bm z\bm z^T\bm]$ are chosen to be respectively consistent estimators of ${\sigma}^{2}$, ${\bm\theta}$ and $E[\bm z\bm z^T\bm]$ computed on ${\cal D}_{k+1}$. The following theorem states the details.

\begin{theorem}\label{th33}
	If $E[{\bm x}]={\bf 0}$, $E[{\bm z}]={\bf 0}$, matrices $E[{\bm x\bm x}^T]$ and $E[{\bm z\bm z}^T]$ are positive definite, $\epsilon\sim N(0,\sigma^2)$, and ${\sigma}_0^{2}$, ${\bm\theta}_0$  and $E_0[\bm z\bm z^T\bm]$ are chosen to be consistent estimators of ${\sigma}^{2}$, ${\bm\theta}$ and $E[\bm z\bm z^T\bm]$ respectively, then, under $H$,
\begin{eqnarray*}&&F_j\stackrel{d}\longrightarrow F_{q,N_j-q} \ \mbox{ in Case 1,} \\&& F^c_j\stackrel{d}\longrightarrow F_{q,N_j-q} \ \mbox{ in both Case 1 and Case 2},\end{eqnarray*} where $F_{q,N_j-q}$ denotes the $F$-distribution on $q$ and $N_j-q$ degrees of freedom.
\end{theorem}

The theorem is the foundation for calculating the quantiles of the test $H:\bm\theta=\bm 0$ given in (\ref{(test-1)}).

\setcounter{equation}{0}
\section{Simulation studies}


In this section, some simulation studies are given to illustrate the effectiveness of our method. For comprehensive comparisons, we take the following three methods as the competitors: the bias-corrected estimate method denoted as BCE, the naive updating estimate method denoted by NUE, and the average estimate method denoted as AVE. Here, the bias-corrected estimate was proposed by Wang et al. (2018); the NUE is defined as the estimator that only makes update using the data segment in which the model keeps unchanged, more specifically, for $j\leq k$,
$$\widetilde{\bm\beta}_j=\left(\bm V_{j-1}^X+\bm X_j^T\bm X_j\right)^{-1}\left(\bm V_{j-1}^{X}\widetilde{\bm\beta}^0_{j-1}+\bm X_j^T\bm X_j\widehat{\bm\beta}_{n_j}\right),$$
and for $j>k$,
\begin{eqnarray*}
&&\hspace{-8ex}\left(\begin{array}{ll}\widetilde{{\bm\beta}}_j
\\\widetilde{\bm\theta}_j\end{array}\right)
=\left(\begin{array}{llll}{\bm V}^X_{k+1,j-1}+{\bm X}_j^T{\bm X}_j& {\bm V}^{XZ}_{k+1,j-1}+{\bm X}_j^T{\bm Z}_j\\ {\bm V}^{ZX}_{k+1,j-1}+{\bm Z}_j^T{\bm X}_j&{\bm V}^{Z}_{k+1,j-1}+{\bm Z}_j^T{\bm Z}_j
\end{array}\right)^{-1}\\
&&\hspace{-4ex}\nonumber\times
\left(
\left(\begin{array}{llll}{\bm V}^X_{k+1,j-1}& {\bm V}^{XZ}_{k+1,j-1}\\ {\bm V}^{ZX}_{k+1,j-1}&{\bm V}^{Z}_{k+1,j-1}
\end{array}\right)
\left(\begin{array}{ll}\widetilde{\bm\beta}_{j-1}\\ \widetilde{\bm\theta}_{j-1}\end{array}\right)
+\left(\begin{array}{llll}{\bm X}_j^T{\bm X}_j& {\bm X}_j^T{\bm Z}_j\\ {\bm Z}_j^T{\bm X}_j& {\bm Z}_j^T{\bm Z}_j
\end{array}\right)
\left(\begin{array}{ll}\widehat{{\bm\beta}}_{n_j}\\
\widehat{\bm\theta}_{n_j}\end{array}\right)\right);\end{eqnarray*}
and the AVE is defined as estimator that makes average over all local estimators, namely, $\widetilde{\bm\beta}_j=j^{-1}\sum_{i=1}^j\widehat{\bm\beta}_{n_l}$ for $j\leq k$ and $\widetilde{\bm\beta}_j=(j-k)^{-1}\sum_{l=k+1}^j\widehat{\bm\beta}_{n_l}$ for $j> k$. For convenience, we call our method as adaptive updating estimate method, denoted by AUE, as our method is adaptive to the change of covariate set. To evaluate the accuracy of the parameter estimation, the bias and MSE of an estimate based on $R$ simulations are reported. For example, the bias and MSE of $\widetilde{\bm\beta}_j$ are respectively defined as $\mbox{bias}(\widetilde{\bm\beta}_j)=|\frac{1}{R}\sum_{i=1}^R(\widetilde{\bm\beta}_j^{(i)}-\bm\beta)|/p$ and $\mbox{MSE}(\widetilde{\bm\beta}_j)=\frac{1}{R}\sum_{i=1}^R\|\widetilde{\bm\beta}_j^{(i)}-\bm\beta\|^2/p$, where $\widetilde{\bm\beta}_j^{(i)}$ the estimate of $\widetilde{\bm\beta}_j$ in the $i$-th simulation.

{\bf Example 1.} We first consider the simple case where there is only once update of model across the whole updating period. The true model behind data is formulated as
\begin{equation*}
y = \bm x^T\bm\beta+\bm z^T\bm\theta+\varepsilon,
\end{equation*}
where $\varepsilon\sim N(0,2)$ is random error, $\bm x=(X_1,\cdots,X_p)$ and $\bm z=(X_{p+1},\cdots,X_{p+q})$, $X_j$ for $j=1,\cdots,p+q$ follow the standard normal distribution with covariance matrix $\bm\Sigma=(\sigma_{ij})_{1\leq i,j\leq p+q}$. We consider two different settings for $\sigma_{ij}$.
\begin{itemize}
	\item [Case] 1.
The covariates $\bm x$ and $\bm z$ are uncorrelated: $\sigma_{ij}=0$ for $i\in\{1,\cdots,p\}$ and $j\in\{p+1,\cdots,p+q\}$, but $\sigma_{ij}=0.5^{|i-j|}$ for $i,j\in\{1,\cdots,p\}$ and $i,j\in\{p+1,\cdots,p+q\}$.
\item [Case] 2. The covariates $\bm x$ and $\bm z$ are correlated: $\sigma_{ij}=0.5^{|i-j|}$ for $1\leq i,j\leq p+q$.
\end{itemize}
The model parameters are set as follows:
\begin{itemize}
	\item [(a)] $\bm\beta=(1,-1)^T$ and  $\bm\theta=1$.
	\item [(b)] $\bm\beta=(1,-1,2,-0.5,0.5)^T$ and  $\bm\theta=(1,-1)$.
\end{itemize}
Let $(\bm y_l,\bm X_l, \bm Z_l)$ be the $l$-th data stream from the population $(y,\bm x,\bm z)$, but note that before the $k$-th data stream, we can only observe data set $(\bm y_l,\bm X_l)$ with $l\leq k$.
The augmentation of $\bm z$ occurs from the 11-th data stream on. For simplicity, we fix the sample size of each data batch as $n$.
Table \ref{tab_ex1_1}-\ref{tab_ex1_2} presents the bias and MSE of $\widetilde{\bm\beta}_j$ and $\widetilde{\bm\theta}_j$ along with the updating procedure. From the tables, the following conclusions can be observed:
\begin{enumerate}
	\item When $\bm x$ and $\bm z$ are uncorrelated, our method has a superior performance over all the competitors. More precisely, from the beginning of the covariate $\bm z$ being available, namely, $j\geq 12$, it can be seen that the bias and MSE of AUE are significantly smaller that those of the competitors.
	\item When $\bm x$ and $\bm z$ get correlated, the results are a little different in the sense that the AUE behaves comparably with BCE at stage $j=12$ but significantly outperforms the other methods when $j$ gets larger. The reason might be that, at the beginning, the estimate of $\bm B$ is not very accurate but when more data batches arrive, $\bm B$ can be estimated very well, so the corresponding Bias and MSE of AUE of $\bm\beta$ are reduced gradually.
	\item For the estimate of $\bm\theta$, it can be observed that AUE, BCE and NUE have a comparable performance and all of them behave better than AVE.
	\item A larger sample size without doubt results in a more accurate simulation result, implying the consistency of AUE.
\end{enumerate}

{\bf Example 2.} In this example, our goal is to examine the adaptability to the case where the covariate $\bm z$ is in fac not active to the response $y$ but it is added into the model when it is available, i.e., to check  whether the estimation of $\bm\beta$ will be affected by setting $\bm\theta=\bm0$. We continue to use the model in Example 1 as an illustration but only consider the case  (b). The simulation results are presented in Table \ref{tab_ex2}. We can observe the very similar results to those in Example 1, which implies that the effectiveness of $\widetilde{\bm\beta}_j$ will not be affected even we take the null predictor $\bm z$ into consideration.

\begin{table}[H]
\centering \small
\caption{Simulation results of Example 1: Bias and MSE of $\widetilde{\bm\beta}_j$($\times 10^{-2}$)}\label{tab_ex1_1}
\begin{tabular}{ccccccccccc} \hline
	&		&		&\multicolumn{4}{c}{Case (a)}&\multicolumn{4}{c}{Case (b)}\\ \cline{4-11}
$n$	&$j$  	&    	&AUE   &BCE  &NUE &AVE&AUE   &BCE  &NUE&AVE \\ \hline
&\multicolumn{10}{c}{$\bm x$ and $\bm z$ are uncorrelated}\\
50 	&$12$	&Bias	&0.2162&0.2242&0.5332&0.5859&0.1266&0.2298&0.2484&0.2256 \\
	&		&MSE		&0.4603&1.4910&2.1853&2.2982&0.5817&1.7266&2.1842&2.4272 \\
	&$16$	&Bias	&0.1814&0.1430&0.1957&0.1839&0.1211&0.1228&0.2265&0.1869 \\
	&		&MSE		&0.3256&0.6098&0.7047&0.7427&0.3595&0.6025&0.6621&0.7881 \\
	&$20$	&Bias	&0.0855&0.1777&0.2740&0.2311&0.1005&0.1613&0.0690&0.0617 \\
	&		&MSE		&0.2439&0.3712&0.3975&0.4300&0.2683&0.3778&0.3994&0.4776 \\
100	&$12$	&Bias	&0.1379&0.1737&0.2074&0.1832&0.1226&0.2168&0.2639&0.3199 \\
	&		&MSE		&0.2249&0.6913&1.0719&1.0894&0.2810&0.7790&1.0591&1.1041 \\
	&$16$	&Bias	&0.1006&0.1346&0.0637&0.0737&0.1301&0.1538&0.1650&0.2053 \\
	&		&MSE		&0.1557&0.2901&0.3254&0.3391&0.1850&0.3162&0.3476&0.3747 \\
	&$20$	&Bias	&0.0854&0.1212&0.1188&0.1140&0.1160&0.0922&0.1293&0.1458 \\
	&		&MSE	&0.1175&0.1812&0.1953&0.2059&0.1321&0.1906&0.2072&0.2234 \\
&\multicolumn{10}{c}{$\bm x$ and $\bm z$ are correlated}\\
50	&$12$	&Bias	&0.6697&0.4467&0.5399&0.5551&0.4294&0.2792&0.3561&0.3525 \\
	&		&MSE		&2.4094&2.2121&3.2892&3.4322&3.1204&2.7786&3.8206&4.2425 \\
	&$16$	&Bias	&0.2991&0.3677&0.2955&0.2905&0.2735&0.1994&0.2131&0.1899 \\
	&		&MSE		&0.7621&0.9191&1.0790&1.1284&0.8322&1.0310&1.1606&1.3902 \\
	&$20$	&Bias	&0.1492&0.2625&0.2896&0.2616&0.1691&0.1876&0.1096&0.0797 \\
	&		&MSE		&0.4540&0.5656&0.5921&0.6312&0.5172&0.6618&0.7061&0.8481 \\
100	&$12$	&Bias	&0.2088&0.2434&0.2443&0.2178&0.2137&0.2598&0.3740&0.4337 \\
	&		&MSE		&1.1185&1.0227&1.6280&1.6497&1.2341&1.2509&1.8655&1.9484 \\
	&$16$	&Bias	&0.1712&0.2902&0.0851&0.0937&0.1290&0.1978&0.2462&0.2919 \\
	&		&MSE		&0.3542&0.4336&0.4996&0.5223&0.4080&0.5452&0.6098&0.6554 \\
	&$20$	&Bias	&0.0666&0.1377&0.1375&0.1230&0.1920&0.1376&0.1808&0.2025 \\
	&		&MSE		&0.2229&0.2703&0.2985&0.3153&0.2612&0.3308&0.3562&0.3842 \\  \hline
\end{tabular}
\end{table}

\begin{table}[H]
\centering\small
\caption{Simulation results of Example 1: Bias and MSE of $\widetilde{\bm\theta}_j$($\times 10^{-2}$)}\label{tab_ex1_2}
\begin{tabular}{ccccccccccc} \hline
&		&		&\multicolumn{4}{c}{Case (a)}&\multicolumn{4}{c}{Case (b)}\\ \cline{4-11}
	$n$	&$j$  	&    	&AUE   &BCE  &NUE &AVE&AUE   &BCE  &NUE&AVE \\ \hline
&\multicolumn{10}{c}{$\bm x$ and $\bm z$ are uncorrelated}\\
50	&12	&Bias&0.2610&0.5042&0.2484&0.2440&0.2315&1.4311&0.1909&0.0225\\
	&	&MSE &2.0384&2.0387&1.9636&2.0441&2.1299&2.1631&2.2048&2.4081\\
	&16	&Bias&0.2721&0.5838&0.2781&0.3391&0.2802&0.7487&0.0532&0.1260\\
	&	&MSE &0.6828&0.6861&0.6592&0.7004&0.6727&0.6740&0.6822&0.8018\\
	&20	&Bias&0.0385&0.2621&0.0762&0.1419&0.0678&0.4344&0.0297&0.0474\\
	&	&MSE &0.4072&0.4082&0.3755&0.3932&0.3795&0.3829&0.3821&0.4572\\
100	&12	&Bias&0.4508&0.7292&0.1601&0.1757&0.2193&0.7996&0.2276&0.1889\\
	&	&MSE &1.0752&1.0824&1.1432&1.1841&1.0851&1.0923&1.0621&1.1150\\
	&16	&Bias&0.1912&0.3315&0.0796&0.0768&0.1912&0.3315&0.0796&0.0768\\
	&	&MSE &0.3405&0.3408&0.3627&0.3733&0.3405&0.3408&0.3627&0.3733\\
	&20	&Bias&0.0612&0.0456&0.0962&0.0563&0.0612&0.0456&0.0962&0.0563\\
	&	&MSE &0.2026&0.2032&0.2149&0.2248&0.2026&0.2032&0.2149&0.2248\\
&\multicolumn{10}{c}{$\bm x$ and $\bm z$ are correlated}\\
50	&12	&Bias&0.2747&0.5067&0.3043&0.2988&0.3147&1.7516&0.2323&0.0275\\
	&	&MSE &3.0575&3.0521&2.9453&3.0662&3.7442&3.7840&3.8780&4.2243\\
	&16	&Bias&0.3332&0.6526&0.3406&0.4153&0.3515&0.9078&0.0692&0.1511\\
	&	&MSE &1.0242&1.0256&0.9889&1.0506&1.1704&1.1680&1.1982&1.4072\\
	&20	&Bias&0.0471&0.2824&0.0933&0.1738&0.0806&0.5440&0.0421&0.0548\\
	&	&MSE &0.6108&0.6113&0.5632&0.5898&0.6669&0.6724&0.6694&0.8001\\
100	&12	&Bias&0.5521&0.8325&0.1961&0.2152&0.2956&1.0220&0.2857&0.2386\\
	&	&MSE &1.6128&1.6203&1.7148&1.7761&1.8791&1.8867&1.8766&1.9693\\
	&16	&Bias&0.2341&0.3747&0.0975&0.0941&0.2531&0.0860&0.3196&0.3226\\
	&	&MSE &0.5108&0.5108&0.5440&0.5600&0.5884&0.5899&0.6049&0.6446\\
	&20	&Bias&0.0749&0.0365&0.1178&0.0689&0.0628&0.2763&0.2114&0.2050\\
	&	&MSE &0.3038&0.3044&0.3223&0.3372&0.3485&0.3488&0.3410&0.3694\\
\hline
\end{tabular}
\end{table}

\begin{table}[H]
\centering\small
\caption{Simulation results of Example 2: Bias and MSE of $\widetilde{\bm\beta}_j$ and $\widetilde{\bm\theta}_j$ ($\times 10^{-2}$)}\label{tab_ex2}
\begin{tabular}{ccccccccccc} \hline
&	&	&\multicolumn{4}{c}{Case (a)}&\multicolumn{4}{c}{Case (b)}\\ \cline{4-11}
&$j$ &    	&AUE   &BCE  &NUE &AVE&AUE   &BCE  &NUE&AVE \\ \hline
&&\multicolumn{9}{c}{$\bm x$ and $\bm z$ are uncorrelated}\\
$\widetilde{\bm\beta}_j$
&12	&Bias&0.1297&0.1891&0.2074&0.1832&0.0676&0.1517&0.2639&0.3199\\
&	&MSE &0.1591&0.5463&1.0719&1.0894&0.1629&0.5560&1.0591&1.1041\\
&16 &Bias&0.0764&0.1256&0.0637&0.0737&0.0555&0.1300&0.1650&0.2053\\
&	&MSE &0.1246&0.2703&0.3254&0.3391&0.1242&0.2895&0.3476&0.3747\\
&20 &Bias&0.0461&0.1132&0.1188&0.1140&0.0585&0.1073&0.1293&0.1458\\
&	&MSE &0.0995&0.1762&0.1953&0.2059&0.0970&0.1817&0.2072&0.2234\\
&	&\multicolumn{9}{c}{$\bm x$ and $\bm z$ are correlated}\\
&12 &Bias&0.3323&0.3181&0.2443&0.2178&0.1932&0.2082&0.3740&0.4337\\
&	&MSE &0.4679&0.8766&1.6280&1.6497&0.5307&1.0172&1.8655&1.9484\\
&16 &Bias&0.0876&0.1993&0.0851&0.0937&0.0659&0.1755&0.2462&0.2919\\
&	&MSE &0.2372&0.4146&0.4996&0.5223&0.2582&0.5164&0.6098&0.6554\\
&20 &Bias&0.0388&0.1288&0.1375&0.1230&0.0702&0.1504&0.1808&0.2025\\
&	&MSE &0.1759&0.2639&0.2985&0.3153&0.1915&0.3201&0.3562&0.3842\\
&	&\multicolumn{9}{c}{$\bm x$ and $\bm z$ are uncorrelated}\\
$\widetilde{\bm\theta}_j$
&12	&Bias&0.4508&0.4169&0.1601&0.1757&0.2193&0.2295&0.2276&0.1889\\
&	&MSE &1.0752&1.0740&1.1432&1.1841&1.0851&1.0749&1.0621&1.1150\\
&16	&Bias&0.1912&0.1621&0.0796&0.0768&0.3238&0.3297&0.2631&0.2626\\
&	&MSE &0.3405&0.3391&0.3627&0.3733&0.3361&0.3360&0.3432&0.3659\\
&20	&Bias&0.0612&0.0579&0.0962&0.0563&0.0520&0.0481&0.1779&0.1712\\
&	&MSE &0.2026&0.2022&0.2149&0.2248&0.2006&0.1996&0.1940&0.2103\\
&	&\multicolumn{9}{c}{$\bm x$ and $\bm z$ are correlated}\\
&12	&Bias&0.5521&0.5158&0.1961&0.2152&0.2956&0.3101&0.2857&0.2386\\
&	&MSE &1.6128&1.6096&1.7148&1.7761&1.8791&1.8623&1.8766&1.9693\\
&16	&Bias&0.2341&0.2003&0.0975&0.0941&0.4512&0.4611&0.3196&0.3226\\
&	&MSE &0.5108&0.5088&0.5440&0.5600&0.5884&0.5889&0.6049&0.6446\\
&20	&Bias&0.0749&0.0717&0.1178&0.0689&0.0628&0.0606&0.2114&0.2050\\
&	&MSE &0.3038&0.3033&0.3223&0.3372&0.3485&0.3467&0.3410&0.3694\\
\hline
\end{tabular}
\end{table}

\begin{figure}[H]
	\centering\small
	\includegraphics[width=6in]{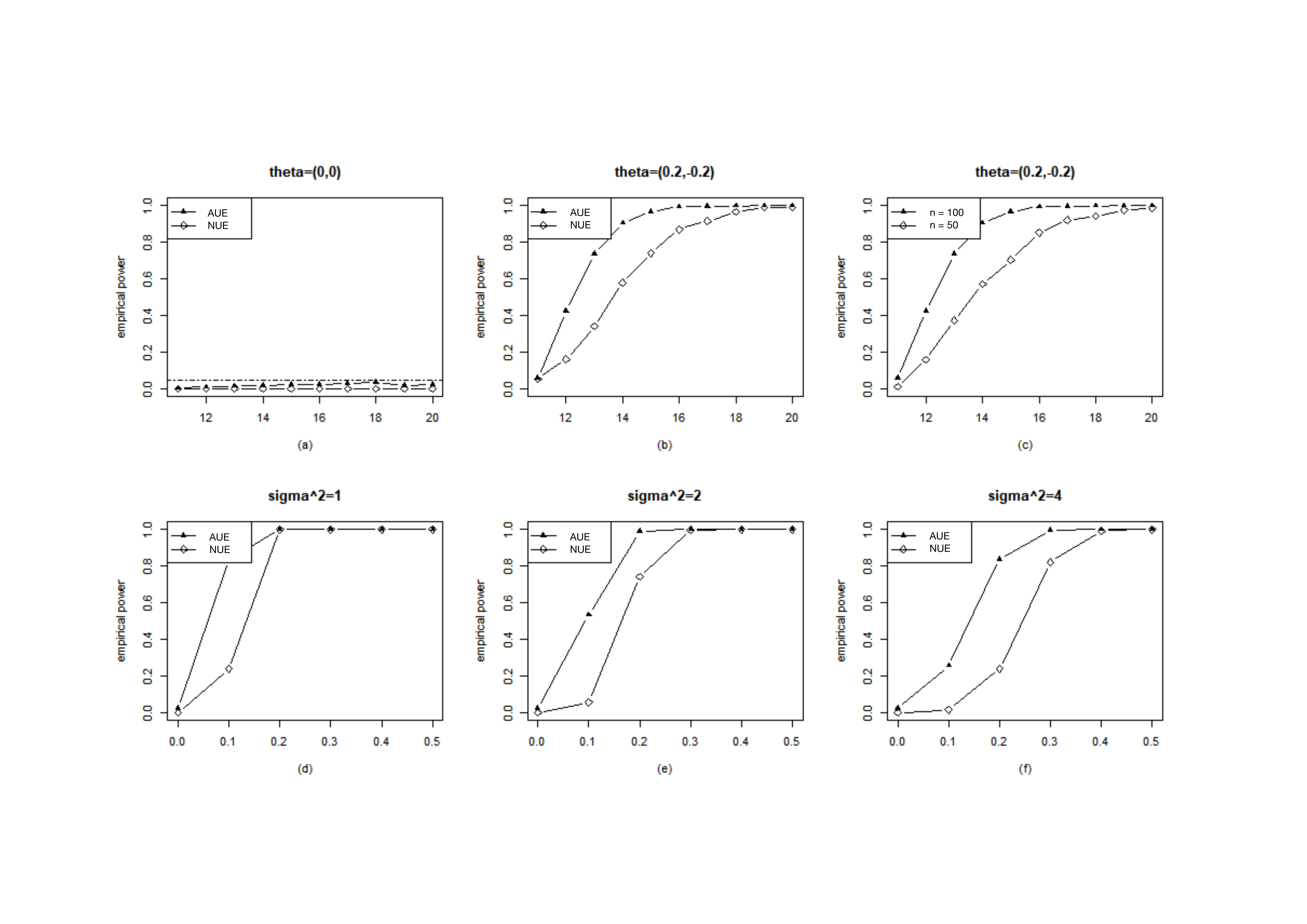}
	\caption{The empirical power curves of the methods AUE and NUE under different model settings.}\label{fig_ep}
\end{figure}

{\bf Example 3.} This example is used to check the effectiveness of the $F$-statistic in the hypothesis of testing $\bm\theta=\bm 0$. We use the model in Example 1 by setting $\bm\beta=(1,-1,2,-0.5,0.5)^T$ and $\bm\theta=a(1,-1)^T$. The predictors are drawn from standard normal distribution with the same condition as in Case 2 of Example 1. We consider the following different model settings:
\begin{enumerate}
	\item $a=0$ with $n=100$. Under this situation, the plots of the empirical power of AUE and NUE respectively against $j$ are reported in Figure \ref{fig_ep}(a).
	\item $a=1$ with $n=100$. Under this situation, the plots of the empirical power of AUE and NUE respectively against $j$ are reported in Figure \ref{fig_ep}(b).
	\item $a=1$ with $n=50$ and $n=100$. Under the two situations, the plots of the empirical power of AUE against $j$ are presented in Figure \ref{fig_ep}(c).
	\item $a=1$ with $n=100$ and different error variances.
Under these situations, the plots of the empirical power of AUE and NUE respectively against $a$ are given in Figures \ref{fig_ep} (d)-(f).
\end{enumerate}
From these figures, the following results can be summarized. First of all, from Figure (a), it can be seen that the size of both AUE and NUE can be controlled well below 0.05. From Figure (b) we can see that  the test statistic of AUE is much more powerful than that of NUE. Figure (c) indicates that a larger sample size $n=100$ improves the power of the testing statistics. Finally, from Figures (d)-(f), we still can observe that AUE performs better than NUE and, although the large variance of random error does affect the performance of the testing statistics, the incidence is controlled within certain limits.

\newpage

{\bf Example 4.} In this example, we consider twice updating for the linear model. Suppose the data are from the following model:
\begin{equation*}
	Y=\bm x^T\bm\beta +\bm z^T\bm\theta+\bm w^T\bm\gamma+\varepsilon.
\end{equation*}
All predictors follow the standard normal distribution with the same setting as Case 2 in Example 1, the parameters are set as $\bm\beta=(1,-1,0.5,-0.5)^T$, $\bm\theta=(1,-1,0.5)^T$ and $\bm\gamma=(1,-0.5)^T$, and $\varepsilon\sim N(0,\sigma^2)$. Before the $11$-th data batch, only the predictor $\bm x$ is available, from the beginning of 11-th data batch, the covariate $\bm z$ is added into the model, finally, after the 21-th data batch, the covariate $\bm w$ is observed. We fix the sample size of data batch as 100 and consider two values for $\sigma^2$ equal to 2 and 4, respectively. The simulation results are shown in Table \ref{tab_ex4}. In this example, we only compare our method with NUE and AVE as Wang et al. (2018) does not provide the updating formula for the case that the model is updated twice.  From this table, it can be found that both the estimate of $\bm\beta$ and $\bm\theta$ are improved significantly by our method, and our estimator of $\bm\gamma$ is not bad compared with the other two methods.

\begin{table}[H]
	\centering
	\caption{Simulation results of Example 4: Bias and MSE of $\widetilde{\bm\beta}_j$, $\widetilde{\bm\theta}_j$ and $\widetilde{\bm\gamma}_j$ ($\times 10^{-2}$)}\label{tab_ex4}
	\begin{tabular}{ccccccccc}
	\hline
	&&&\multicolumn{3}{c}{$\sigma^2=2$}&\multicolumn{3}{c}{$\sigma^2=4$}\\ \cline{4-9}
	$j$&&&AUE&NUE &AVE&AUE&NUE&AVE\\ \hline
	25	&$\bm\beta$  &Bias&0.3468&0.3142&0.2826&0.3989&0.4443&0.3996\\
		&			 &MSE &0.6043&0.7169&0.7913&0.8270&1.4339&1.5826\\
		&$\bm\theta$ &Bias&0.2331&0.1878&0.1655&0.2817&0.2656&0.2341\\
		&			 &MSE &0.5173&0.7299&0.7967&0.8027&1.4597&1.5934\\
		&$\bm\gamma$ &Bias&0.1951&0.0088&0.0942&0.2759&0.0124&0.1332\\
		&			 &MSE &0.7449&0.7387&0.8191&1.4899&1.4774&1.6382\\
	30	&$\bm\beta$  &Bias&0.2277&0.1439&0.1090&0.3307&0.2036&0.1541\\
		&			 &MSE &0.3379&0.3537&0.3883&0.4956&0.7074&0.7765\\
		&$\bm\theta$ &Bias&0.1491&0.0395&0.0194&0.1799&0.0559&0.0274\\
		&		 	 &MSE &0.2869&0.3496&0.3825&0.4839&0.6993&0.7650\\
		&$\bm\gamma$ &Bias&0.0714&0.0612&0.0290&0.1009&0.0866&0.0410\\
		&		 	 &MSE &0.3554&0.3419&0.3892&0.7108&0.6838&0.7783\\
	\hline
	\end{tabular}
\end{table}

\section{Conclusions and future works}

It was shown in Introduction that although a large number of statistical methods and computational recipes have been developed to address the challenge of analyzing the models in data streams, the models with varying covariate set are rarely investigated in the existing literature. Furthermore, simple and unified strategy has not been introduced, the relevant oracle properties have not been established. In the previous sections of this paper, a homogenization technique was introduced, and then a unified online updating strategy was proposed to consistently and efficiently estimate the parameters, and further to establish various statistics such as residual sum of squares and $F$-statistic in updating regression models. The newly proposed method is computational simple and achieves the oracle properties. These properties are always satisfied whenever the model updating happens, without any constraint on the number of data batches.

The behavior of the method was further illustrated by various
numerical examples from simulation experiments. The simulation verified that the
finite performance of the new method is much better than the competitors such as simple average, bias-correction method and naive estimator, and the new estimator has
the similar behavior as that of the oracle estimator obtained by using simultaneously the entire data sets, rather than sequent updating.

However, in this paper we focus only on the linear regression models in data streams.
It is still a challenge to extend the homogenization strategy to more general models such as general nonlinear models and nonparametric models. This is an interesting issue and is worth further study in the future.

\section*{References}
\begin{description}

\item Amari, S. I., Park, H. and Fukumizu, K. (2000). Adaptive method of realizing natural gradient learning for multilayer perceptrons. {\it Neural Computation}, {\bf 12}, 1399-1409.

\item Bordes, A., Bottou, L. and Gallinari, P. (2009). Sgd-qn: careful quasi-Newton stochastic gradient descent. {\it Journal of Machine Learning Research}, {\bf 10}, 1737-1754.

\item Bucak, S. S. and Gunsel, B. (2009). Incremental subspace learning via non-negative matrix factorization. {\it Pattern Recogination}, {\bf 42}, 788-797.

    \item Cai, Z. R., Li, R. Z. and Zhu, L. P. (2020).  Online sufficient dimension reduction through sliced
inverse regression. {\it Journal of Machine Learning Research}, {\bf 21}, 1-25.

\item Certo, S. T. (2003). Influencing initial public offering investors with prestige: Signaling with board structures. {\it Academy of Management Review}, {\bf 28}, 432-446.

\item Chen, X. and Xie, M. G. (2014). A split-and-conquer approach for analysis of extraordinarily large data. {\it Statistica Sinica}, {\bf 24}, 1655-1684.

\item Desyllas, P. and Sako, M. (2013). Profiting from business model innovation: Evidence from pay-as-you-drive auto insurance. {\it Research Policy}, {\bf 42}, 101-116.

\item Duchi, J., Hazan, E. and Singer, Y. (2011). Adaptive subgradient methods for online learning and stochastic optimization. {\it Journal of Machine Learning Research}, {\bf 12}, 2121-2159.


\item Hazan, E., Agarwal, A. andKale, S. (2007). Logarithmic regret algorithms for online convex optimization. {\it Journal of Machine Learning Research}, {\bf 69}, 169-192.

\item Hao, S., Zhao, P., Lu, J., Hoi, S. C. H., Miao, C. and Zhang, C. (2016). Soal: second-order online active learning. {\it International Conference of Data Mining, Barcelona}.

\item Kleiner, A., Talwalkar, A., Sarkar, P. and Jordan, M. I. (2014) A scalable bootstrap for massive data. {\it Journal of The Royal Statistical Society Series B}, {\bf 76}, 795-816.


\item Li, K. and Yang, J. (2018). Score-Matching Representative Approach for Big Data
Analysis with Generalized Linear Models. \textit{arXiv}: 1811.00462.

\item Li, R., Lin, D. K. and Li, B. (2013). Statistical inference in massive data sets. {\it Applied
Stochastic Models in Business and Industry}, {\bf 29}, 399-409.

\item Liang, F., Cheng, Y., Song, Q., Park, J. and Yang, P. (2013). A resampling-based stochastic approximation method
for analysis of large geostatistical data. {\it Journal of American Statistical Association}, {\bf 108}, 325-339.

\item Lin, N. and  Xi, R. (2011). Aggregated estimating equation estimation. {\it Statistics and Its Interface}, {\bf 4}, 73-83.


\item Liu, D. C. and Nocedal, J. (1989). On the limited memory bfgs method for large scale optimization.
{\it Mathematical Programming}, {\bf 45}, 503-528.

\item Luo, L. and Song, P. X. K. (2020). Renewable estimation and incremental inference in
generalized linear models with streaming data sets. {\it Journal of The Royal Statistical Society Series B}, {\bf 82}, 69-97.

\item Ma, P., Mahoney, M.W. and Yu, B. (2015). A statistical perspective on algorithm leveraging. {\it Journal of Machine Learning Research}, {\bf 6}, 861-911.

\item McCulloch, C. and Searle, S. (2001). {\it Generalized, Linear, and mixed models.} Hohn Wiley and Sons, Inc. New Work.

\item Maclaurin, D.and Adams, R. P. (2014). Firefly Monte Carlo: Exact MCMC with subsets of data. \textit{arXiv}:1403.5693.


\item Neiswanger,W.,Wang, C., and Xing, E. (2013). Asymptotically exact, embarrassingly parallel MCMC. preprint. \textit{arXiv}:1311.4780.

\item Nion, D. and Sidiropoulos, N. D. (2009). Adaptive algorithms to track the PARAFAC decomposition of a thirdorder tensor. {\it IEEE Transaction on Signal Process}, {\bf 57}, 2299-2310.


\item Nocedal, J. and Wright, S. J. (1999). {\it Numerical optimization}. Springer-Verlag, New York.

\item Pillonetto, G., Schenato, L. and Varagnolo, D. (2019). Distributed multi-agent gaussian regression
via finite-dimensional approximations. {\it IEEE Transactions on Pattern Analysis and Machine Intelligence,} {\bf 41}, No. 9, 2098-2111.

\item Robbins, H. and Monro, S. (1951). A stochastic approximation method. {\it Annals of Mathematical Statistics}, {\bf 22}, 400-407.

\item Schifano, E. D., Wu, J., Wang, C., Yan, J. and Chen, M. H. (2016). Online updating of statistical inference in the
big data setting. {\it Technometrics}, {\bf 58}, 393-403.

\item Scott, S. L., Blocker, A. W., Bonassi, F. V., Chipman, H., George, E., and  McCulloch, R. (2013). {\it Bayes and Big Data: The Consensus Monte Carlo Algorithm, EFaBBayes 250 Conference}, 16.

\item Schraudolph, N. N., Yu, J. and G\"{u}nter, S. (2007).  A stochastic quasi-Newton method for online convex optimization. {\it Proceedings of Machine Learning Research}, {\bf 2}, 436-443.

\item Song, Q. and Liang, F. (2014). A split-and-merge Bayesian variable selection approach for ultrahigh dimensional regression. {\it Journal of the Royal Statistical Society Series B}, {\bf 77}, 947-972.

\item Toulis, P. and Airoldi, E. M. (2015). Scalable estimation strategies based on stochastic approximations: classical results and new insights. {\it Statistic Computing}, {\bf 25}, 781-795.

\item Vaits, N., Moroshko, E. and Crammer, K. (2015). Second-order non-stationary online learning for regression. {\it Journal of Machine Learning Research}, {\bf 16}, 1481-1517.

\item Wang, C., Chen, M. H., Wu, J., Yan, J., Zhang, Y. and
Schifan, E. (2018). Online updating method with new variables for big data streams. {\it The Canadian Journal of Statistics}, {\bf 46}, 123-146.

\item Wang, H., Yang, M. and Stufken, J. (2019). Information-based optimal subdata selection
for big data linear regression. {\it Journal of the American Statistical Association},
{\bf 114}, 393-405.

\item Wang, S. G., Wu, M. X and Jia, Z. Z. (2006). {\it Matrix inequalities}. Science Press, Beijing.


\item Xue, Y.,  Wang, H., Yan, J. and Schifano, E. D. (2019). An online updating approach for testing the proportional
hazards assumption with streams of survival data. {\it Biometric} (to appear).

\end{description}

\setcounter{equation}{0}
\section{Appendix: Proofs}

{\it Proof of Theorem
\ref{th31}.}
By empirical average models in (\ref{(eq-4)}), we have
\begin{eqnarray*}
\left(\begin{array}{ll}\widetilde{{\bm\beta}}_j\\\widetilde{\bm\theta}_j\end{array}\right)
&=&\left(\begin{array}{cccc}{\bm V}^X_j&{\bm V}^{XZ}_{k+1,j}\\ {\bm V}^{ZX}_{k+1,j}&{\bm V}^{Z}_{k+1,j}
\end{array}\right)^{-1}\left(\begin{array}{ll}{\bm V}^{X}_j\bm\beta +{\bm V}^{XZ}_{k+1,j}{\bm\theta}+{\bm V}^{X\bm\varepsilon}_k+{\bm V}^{X\bm\epsilon}_{k+1,j}
\\  {\bm V}^{ZX}_{k+1,j}\bm\beta + {\bm V}^{Z}_{k+1,j}{\bm\theta}+ {\bm V}^{Z\bm\epsilon}_{k+1,j}\end{array}\right)\\&=&
\left(\begin{array}{ll}{\bm\beta}\\\bm\theta \end{array}\right)+\left(\begin{array}{cccc}{\bm V}^X_j&{\bm V}^{XZ}_{k+1,j}\\ {\bm V}^{ZX}_{k+1,j}&{\bm V}^{Z}_{k+1,j}
\end{array}\right)^{-1}\left(\begin{array}{ll}{\bm V}^{X\bm\varepsilon}_k+{\bm V}^{X\bm\epsilon}_{k+1,j}
\\  {\bm V}^{Z\bm\epsilon}_{k+1,j}\end{array}\right)
\\&=&
\left(\begin{array}{ll}{\bm\beta}\\\bm\theta \end{array}\right)+\left(\begin{array}{cccc}\frac{1}{N_j}{\bm V}^X_j&\frac{1}{N_j}{\bm V}^{XZ}_{k+1,j}\\ \frac{1}{N_j}{\bm V}^{ZX}_{k+1,j}&\frac{1}{N_j}{\bm V}^{Z}_{k+1,j}
\end{array}\right)^{-1}\left(\begin{array}{ll}\frac{1}{N_j}{\bm V}^{XZ}_k\bm\theta+\frac{1}{N_j}{\bm V}^{X\bm\epsilon}_{j}
\\  \frac{1}{N_j}{\bm V}^{Z\bm\epsilon}_{k+1,j}\end{array}\right).\end{eqnarray*}
It is supposed that ${\bm x}$ and ${\bm z}$ are uncorrelated, $E[{\bm z}]=0$ and $E({\bm\epsilon}|{\bm x},{\bm z})=0$. This implies  $$\left(\begin{array}{ll}\frac{1}{\sqrt {N_j}}{\bm V}^{XZ}_k\bm\theta+\frac{1}{\sqrt {N_j}}{\bm V}^{X\bm\epsilon}_{j}
\\  \frac{1}{\sqrt {N_j}}{\bm V}^{Z\bm\epsilon}_{k+1,j}\end{array}\right)\stackrel{d}\longrightarrow
N\left(0, \Phi_\rho \right),$$ where
\begin{eqnarray*}\Phi_\rho&=&\left(\begin{array}{cccc}\frac{1-\rho}{\overline\sigma^2_\varepsilon} E[(\bm\theta^T {\bm z})^2]E[{\bm x\bm x}^T]+\frac{\sigma^2(1-\rho)}{\overline\sigma^2_\varepsilon}  E[{\bm x\bm x}^T]+\frac{\sigma^2\rho}{\sigma^2_0}  E[{\bm x\bm x}^T]&{\bf 0}\\ {\bf 0} &\frac{\sigma^2\rho}{\sigma_0^2} E[{\bm z\bm z}^T]
\end{array}\right)\\&=&\left(\begin{array}{cccc}\frac{\sigma^2_\varepsilon(1-\rho)}{\overline\sigma^2_\varepsilon}  E[{\bm x\bm x}^T]+\frac{\sigma^2\rho}{\sigma^2_0}  E[{\bm x\bm x}^T]&{\bf 0}\\ {\bf 0} &\frac{\sigma^2\rho}{\sigma_0^2} E[{\bm z\bm z}^T]
\end{array}\right).\end{eqnarray*}
Then,
\begin{eqnarray*}&&
\sqrt{N_j}\left(\left(\begin{array}{ll}\widetilde{{\bm\beta}}_j\\\widetilde{\bm\theta}_j\end{array}\right)
-\left(\begin{array}{ll}{\bm\beta}\\\bm\theta \end{array}\right)\right)\\&&=
\left(\begin{array}{cccc}\frac{1}{N_j}{\bm V}^X_j&\frac{1}{N_j}{\bm V}^{XZ}_{k+1,j}\\ \frac{1}{N_j}{\bm V}^{ZX}_{k+1,j}&\frac{1}{N_j}{\bm V}^{Z}_{k+1,j}
\end{array}\right)^{-1}\left(\begin{array}{ll}\frac{1}{\sqrt {N_j}}{\bm V}^{XZ}_k\bm\theta+\frac{1}{\sqrt {N_j}}{\bm V}^{X\bm\epsilon}_{j}
\\  \frac{1}{\sqrt {N_j}}{\bm V}^{Z\bm\epsilon}_{k+1,j}\end{array}\right)\\&&
\stackrel{d}\longrightarrow
N\left(0, \Omega_\rho^{-1}\Phi_\rho \Omega_\rho^{-1}\right),\end{eqnarray*}
where
\begin{eqnarray*}\Omega_\rho=\left(\begin{array}{cccc}\frac{1-\rho}{\overline\sigma_\varepsilon}E[{\bm x\bm x}^T]+\frac{\rho}{\sigma_0}E[{\bm x\bm x}^T]&{\bf 0}\\ {\bf 0}&\frac{\rho}{\sigma_0} E[{\bm z\bm z}^T]
\end{array}\right).\end{eqnarray*}
  $\Box$

 \

\noindent{\it Proof of Corollary \ref{core31} .} It is a direct result of Theorem \ref{th31}.
$\Box$

\vspace{1ex}

\noindent{\it Proof of Theorem \ref{th32}.} Because the estimator $\widehat B$ is consistent, by the formula of the block matrix inversion,
we have
$$\left(\begin{array}{cccc}{\bm V}^X_j&{\bm V}^{X}_{k}\widehat{\bm B}+{\bm V}^{XZ}_{k+1,j}\\ {\bm V}^{ZX}_{k+1,j}&{\bm V}^{Z}_{k+1,j}
\end{array}\right)^{-1}=\left(\begin{array}{cccc}{\bm V}^X_j&{\bm V}^{X}_{k}{\bm B}+{\bm V}^{XZ}_{k+1,j}\\ {\bm V}^{ZX}_{k+1,j}&{\bm V}^{Z}_{k+1,j}
\end{array}\right)^{-1}+o_p(1).$$ Then,
\begin{eqnarray}\label{(proof-1)}
\left(\begin{array}{ll}\widetilde{{\bm\beta}}_j\\\widetilde{\bm\theta}_j\end{array}\right)
=\left(\begin{array}{cccc}{\bm V}^X_j&{\bm V}^{X}_{k}{\bm B}+{\bm V}^{XZ}_{k+1,j}\\ {\bm V}^{ZX}_{k+1,j}&{\bm V}^{Z}_{k+1,j}
\end{array}\right)^{-1}\left(\begin{array}{ll}{\bm V}^{Xy}_j
\\ {\bm V}^{Zy}_{k+1,j}\end{array}\right)+o_p(1)\left(\begin{array}{ll}{\bm V}^{Xy}_j
\\ {\bm V}^{Zy}_{k+1,j}\end{array}\right).\end{eqnarray}
It shows that, asymptotically, $\left(\begin{array}{ll}\widetilde{{\bm\beta}}_j\\\widetilde{\bm\theta}_j\end{array}\right)$ is identically distributed as $$\left(\begin{array}{ll}\widetilde{{\bm\beta}}^*_j\\\widetilde{\bm\theta}^*_j\end{array}\right)=\left(\begin{array}{cccc}{\bm V}^X_j&{\bm V}^{X}_{k}{\bm B}+{\bm V}^{XZ}_{k+1,j}\\ {\bm V}^{ZX}_{k+1,j}&{\bm V}^{Z}_{k+1,j}
\end{array}\right)^{-1}\left(\begin{array}{ll}{\bm V}^{Xy}_j
\\ {\bm V}^{Zy}_{k+1,j}\end{array}\right).$$
By the same argument as used in the proof of Theorem 3.1, we have
\begin{eqnarray*}
\left(\begin{array}{ll}\widetilde{{\bm\beta}}^*_j\\\widetilde{\bm\theta}^*_j\end{array}\right)
=
\left(\begin{array}{ll}{\bm\beta}\\\bm\theta \end{array}\right)+\left(\begin{array}{cccc}\frac{1}{N_j}{\bm V}^X_j&\frac{1}{N_j}{\bm V}^{XZ}_{j}\\ \frac{1}{N_j}{\bm V}^{ZX}_{k+1,j}&\frac{1}{N_j}{\bm V}^{Z}_{k+1,j}
\end{array}\right)^{-1}\left(\begin{array}{ll}\frac{1}{N_j}{\bm V}^{X\bm\epsilon}_{j}
\\  \frac{1}{N_j}{\bm V}^{Z\bm\epsilon}_{k+1,j}\end{array}\right).\end{eqnarray*}
It can be seen that  $$\left(\begin{array}{ll}\frac{1}{\sqrt {N_j}}{\bm V}^{X\bm\epsilon}_{j}
\\  \frac{1}{\sqrt {N_j}}{\bm V}^{Z\bm\epsilon}_{k+1,j}\end{array}\right)\stackrel{d}\longrightarrow
N\left(0, \Phi^c_\rho \right),$$ where
$$\Phi^c_\rho=\left(\begin{array}{cccc}\frac{\sigma_\varepsilon^2(1-\rho)}{\overline\sigma^2_\varepsilon}  E[{\bm x\bm x}^T]+\frac{\sigma^2\rho}{\sigma^2_0} E[{\bm x\bm x}^T]&\frac{\sigma^2\rho}{\sigma^2_0} E[{\bm x\bm z}^T]\\ \frac{\sigma^2\rho}{\sigma_0^2} E[{\bm z\bm x}^T] &\frac{\sigma^2\rho}{\sigma_0^2} E[{\bm z\bm z}^T]
\end{array}\right).$$
Then,
\begin{eqnarray*}&&
\sqrt{N_j}\left(\left(\begin{array}{ll}\widetilde{{\bm\beta}}^*_j\\\widetilde{\bm\theta}^*_j\end{array}\right)
-\left(\begin{array}{ll}{\bm\beta}\\\bm\theta \end{array}\right)\right)=
\left(\begin{array}{cccc}\frac{1}{N_j}{\bm V}^X_j&\frac{1}{N_j}{\bm V}^{XZ}_{j}\\ \frac{1}{N_j}{\bm V}^{ZX}_{k+1,j}&\frac{1}{N_j}{\bm V}^{Z}_{k+1,j}
\end{array}\right)^{-1}\left(\begin{array}{ll}\frac{1}{\sqrt {N_j}}{\bm V}^{X\bm\epsilon}_{j}
\\  \frac{1}{\sqrt {N_j}}{\bm V}^{Z\bm\epsilon}_{k+1,j}\end{array}\right)\\&&
\stackrel{d}\longrightarrow
N\left(0, (\Omega^c_\rho)^{-1}\Phi^c_\rho (\Omega^c_\rho)^{-1}\right),\end{eqnarray*}
where
\begin{eqnarray*}\Omega^c_\rho=\left(\begin{array}{cccc}\frac{1-\rho}{\overline\sigma_\varepsilon}E[{\bm x\bm x}^T]+\frac{\rho}{\sigma_0} E[{\bm x\bm x}^T]&\frac{1-\rho}{\overline\sigma_\varepsilon}E[{\bm z\bm x}^T]+\frac{\rho}{\sigma_0} E[{\bm z\bm x}^T]\vspace{1ex}\\ \frac{\rho}{\sigma_0} E[{\bm x\bm z}^T]&\frac{\rho}{\sigma_0} E[{\bm z\bm z}^T]
\end{array}\right).\end{eqnarray*}
Therefore, by the above result and (\ref{(proof-1)}), we have
\begin{eqnarray*}
\sqrt{N_j}\left(\left(\begin{array}{ll}\widetilde{{\bm\beta}}_j\\\widetilde{\bm\theta}_j\end{array}\right)
-\left(\begin{array}{ll}{\bm\beta}\\\bm\theta \end{array}\right)\right)
\stackrel{d}\longrightarrow
N\left(0, (\Omega^c_\rho)^{-1}\Phi^c_\rho (\Omega^c_\rho)^{-1}\right).\end{eqnarray*}
$\Box$

\vspace{1ex}

\noindent{\it Proof of Theorem \ref{th33}.} Consider the following models
\begin{eqnarray}\label{(proof-2)}\nonumber && \sigma^{-1}_\varepsilon y_{ji}=\sigma^{-1}_\varepsilon {\bm x}_{ji}^T{\bm\beta}+\sigma^{-1}_\varepsilon\varepsilon_{ji},i=1,\cdots,n_j, j>k,\\\label{(eq-5)}&&
{\sigma}^{-1} y_{ji}={\sigma}^{-1}{\bm x}_{ji}^T{\bm\beta}+{\sigma}^{-1}{\bm z}_{ji}^T{\bm\theta}+{\sigma}^{-1}\epsilon_{ji}, i=1,\cdots,n_j, j>k.\end{eqnarray}
The error terms of models in (\ref{(proof-2)}) are identically distribution as $N(0,1)$.
Write
$$F_j|_{w_1=\sigma^{-1}_\varepsilon,w_2=\sigma^{-1}}=\frac{\widetilde{\bm\theta}_j^T\left({\bm V}^{Z}_{k+1,j}-{\bm V}^{ZX}_{k+1,j}({\bm V}^{X}_{j})^{-1}{\bm V}^{XZ}_{k+1,j}\right)\widetilde{\bm\theta}_j/q}{SSE_j/(N_j-q)}\Big|_{w_1=\sigma^{-1}_\varepsilon,w_2=\sigma^{-1}}.$$ Then, $F_j|_{w_1=\sigma^{-1}_\varepsilon,w_2=\sigma^{-1} }$ is the $F$-statistic of  model (\ref{(proof-2)}) for testing $H$. These show that under $H$, $F_j|_{w_1=\sigma^{-1}_\varepsilon,w_2=\sigma^{-1}}\sim F_{q,N_j-q}$.
Moreover, ${\sigma}_0^{2}$, ${\bm\theta}_0$  and $E_0[\bm z\bm z^T\bm]$ are chosen to be consistent estimators of ${\sigma}^{2}$, ${\bm\theta}$ and $E[\bm z\bm z^T\bm]$, respectively.
Then, $$F_j=F_j|_{w_1=\sigma^{-1}_\varepsilon,w_2=\sigma^{-1}}+o_p(1).$$  Therefore, under $H$, $F_j\stackrel{d}\longrightarrow F_{q,N_j-q}$ in Case 1.

By the same argument as used above, we can prove that $F^c_j\stackrel{d}\longrightarrow F_{q,N_j-q}$ in both Case 1 and Case 2.
$\Box$

\end{document}